\documentclass[
 aip,
 amsmath,amssymb,
 reprint,%
]{revtex4-1}

\usepackage{graphicx}
\usepackage{dcolumn}
\usepackage{bm}
\usepackage{mathtools}

\usepackage[utf8]{inputenc}
\usepackage[T1]{fontenc}
\usepackage{mathptmx}
\usepackage{etoolbox}

\usepackage{color}
\usepackage{xcolor}
\usepackage{soul}

\usepackage{multirow, makecell, booktabs, array}
\usepackage[version=3]{mhchem}
\usepackage{boldline} 
\usepackage{siunitx} 

\DeclareSIUnit \angstrom {\text {Å}}
\newcommand{\ag}{\si{\angstrom}}

\makeatletter
\def\@email#1#2{%
 \endgroup
 \patchcmd{\titleblock@produce}
  {\frontmatter@RRAPformat}
  {\frontmatter@RRAPformat{\produce@RRAP{*#1\href{mailto:#2}{#2}}}\frontmatter@RRAPformat}
  {}{}
}%
\makeatother
\begin{document}

\preprint{AIP/123-QED}

\title{Strain-Induced Changes of Electronic and Optical Properties of Zr-based MXenes}
\author{Jiří Kalmár}
\affiliation{Department of Physics, Faculty of Science, University of Ostrava, 30.~dubna 22, 701 03 Ostrava, Czech Republic}
\author{Franti\v{s}ek Karlick\'{y}}
\email{frantisek.karlicky@osu.cz}
\affiliation{Department of Physics, Faculty of Science, University of Ostrava, 30.~dubna 22, 701 03 Ostrava, Czech Republic}


\begin{abstract}
Zr-based MXenes recently attracted attention because of its experimental preparation showing temperature stability, mechanical strength, and promising energy, sensoric, and electrochemistry applications. However, necessary theoretical predictions at a precise/predictive level are complicated due to essential excitonic features and strong electron correlation (i.e., a necessity to go beyond standard density functional theory, DFT). 
Contrary to the prevailing focus on oxygen-terminated MXenes and standard predictions of other Zr-based MXenes as conductors, based on the hybrid DFT and GW many-body perturbational theory, we were able to find seven different semiconductors (five of them for their equilibrium geometry and two others under slight tensile biaxial strain) in the case of two- and three-layered \ce{Zr2CT2} and \ce{Zr3C2T2} configurations with various terminations (T = O, F, S, Cl). 
We observed semiconductor-to-conductor transition induced by strain in the majority of such Zr-based MXenes at experimentally achievable strain range. 
Furthermore, using the Bethe-Salpeter equation (BSE), we demonstrated that selected semiconducting Zr-based MXenes possess high optical absorption efficiency (20-30\%) in the visible light range, underscoring their potential in photonic applications. 
The high sensitivity of Zr-based MXenes to external conditions and functionalization combined with the thermal stability make the materials promising for applications at operational temperatures in electronic and optical technologies.
\\
\end{abstract}

\maketitle

\section{\label{sec:intro}Introduction}
MXenes, a family of recent two-dimensional (2D) carbides, nitrides, and carbonitrides, embody various interesting properties promising for technical applications. \cite{Anasori2017,Hantanasirisakul2018,Gogotsi2019}
Since the first experimentally prepared MXene monolayer, Ti$_3$C$_2$,\cite{Naguib2011} more than 20 MXenes were synthesized, and tens of others were theoretically predicted.\cite{Anasori2020Book, Anasori2022, Lim2022} 
All reported MXenes were terminated by some functional group on the surface (as -O, -OH, -F, or -Cl), MXenes are therefore labeled as M$_n$C$_{n-1}$T$_x$, where M is metal and T is the terminal atom or group. 2D MXene's properties are sensitive to composition, termination, or external conditions (pressure, support, solvent, etc.) and provide a rich set of its phases: metals, semiconductors,\cite{Ketolainen2022} ferromagnets, antiferromagnets\cite{Sakhraoui2022}, topological insulators,\cite{Champagne2021} or excitonic insulators.\cite{Kumar2023}

Here, we are focused on Zr-based MXenes terminated by various groups (O, F, S, Cl) for which the semiconducting behavior could be expected, their electronic and optical properties, and the changes of these properties under biaxial strain. As has been shown recently, the mechanical strain on MXenes can be realized experimentally.\cite{Lipatov2018, Wyatt2021} We note, that Zr-based MXenes are a topic of investigation of a few recent studies\cite{Khazaei2012, Thomas2023, Duan2023, Zhang2024, Zha2015} but these studies are mostly focused on the oxygen-terminated \ce{Zr2CO2} and \ce{Zr3C2O2} MXenes. It is, therefore, our goal to bring a broader understanding of this family of materials and to bring forward less investigated materials with potential applications.

Recently, Zr$_3$C$_2$T$_x$ 2D MXene was experimentally prepared by Zhou et al.\cite{Zhou2016} by a non-traditional method of selectively etching the \ce{Al3C3} layer from nanolaminated \ce{Zr3Al3C5} MAX phase under hydrofluoric acid treatment. The 2D Zr$_3$C$_2$T$_x$ demonstrates an enhanced capability to preserve its two-dimensional characteristics and structural integrity at elevated temperatures in vacuum or argon atmosphere, in comparison to Ti$_3$C$_2$T$_x$ MXenes. 
The difference was explained by a theoretical investigation of binding energy. 
Together with predicted mechanical strengths, it is reasonable to expect that the 2D Zr$_3$C$_2$T$_x$ MXenes will have promising applications from electrical energy storage, reinforcement fillers for polymers, to sensors and catalysts, especially when used in high-temperature environment.\cite{Zhou2016} 
In addition, Zr-MXenes are promising as anode materials of sodium ion batteries.\cite{Zha2022}

The zirconium-based MXenes are often predicted as conductors, with the only exception being the \ce{Zr2CO2} MXene.\cite{Champagne2021, Khazaei2012} A standard tool for the \textit{ab initio} calculations is the density functional theory (DFT), and it has been for years accepted as a reliable and robust computational method. DFT approach can, however, exhibit inconsistencies in its predictions, such as predicting semiconducting materials as gapless,\cite{Yang2016} particularly when applied to complex systems, such as those containing transition metals. These discrepancies often arise due to the different implementation levels of DFT, primarily the diverse density functionals employed.

As has been recently shown, the mechanical biaxial strain can induce indirect to direct semiconductor transitions or can lead to an emergence (or disappearance) of the fundamental band gap in MXenes.\cite{Sakhraoui2022,Zhang2024, Khan2017,Cui2019,Zhang2020} Moreover, due to the change in the band gap, the optical absorption of materials can be red-shifted or blue-shifted when the strain is applied. The possibility of tuning the electronic and optical properties of MXenes thus makes this approach highly promising for the design of efficient (opto)electronic devices.

In this paper, we use state-of-the-art methods to reliably determine the electronic and optical properties of Zr-based MXenes in their natural state and under biaxial strain. Many-body perturbation GW method\cite{Hedin1965} is used to obtain an accurate prediction of the fundamental band gap of semiconducting MXenes and beyond GW approximation the complete comprehension of light absorption is achievable by solving the excitonic equation of motion, known as the Bethe-Salpeter equation (BSE).\cite{Bethe1951} Firstly, we have investigated several possible geometrical conformers of terminated Zr-based MXenes, selected the energetically most favorable ones for corresponding terminal groups, and determined their mechanical stability by analysis of the \textit{ab initio} molecular dynamics (AIMD) simulations. Then, we investigated the changes in band structures under biaxial strain and further selected the most promising conformations. Beyond DFT, we have discovered five semiconducting Zr-based MXenes: \ce{Zr2CO2}, \ce{Zr2CF2}, \ce{Zr2CCl2}, \ce{Zr3C2O2}, \ce{Zr3C2F2}, with two other semiconducting under a relatively small biaxial strain (\ce{Zr2CS2}, \ce{Zr3C2Cl2}). Lastly, we have used advanced many-body methods to determine the fundamental band gap, the optical gap, and the exciton binding energy and their strain-induced changes in these MXenes.

\section{\label{sec:methods}Methods}
Calculations were performed using the periodic density functional theory code Vienna \textit{Ab Initio} Simulation Package (VASP)\cite{kresse_ab_1993,kresse_efficiency_1996,kresse_efficient_1996,kresse_norm-conserving_1994} in version 6.3.1. The DFT Kohn-Sham equations have been solved variationally in a plane-wave basis set using the projector-augmented-wave (PAW) method.\cite{Blochl1994} Structural optimization, ground-state calculations, and band structure calculations were done with Perdew-Burke-Ernzerhof (PBE) density functional\cite{PBE} in generalized gradient approximation (GGA) as well as with meta-GGA Strongly Constrained and Appropriately Normed (SCAN)\cite{SCAN} density functional. Furthermore, for more precise results the wave functions from SCAN functional calculations were used for static calculations using more advanced hybrid density functional HSE06.\cite{HSE06} The convergence criterion for the electronic self-consistency cycle was in all cases set to $10^{-6}$ eV/cell and the structural optimization was stopped if forces were less than $10^{-3}$ eV/$\ag$. The plane-wave cutoff energy was set to 500 eV for all calculations uniformly. GW sets of PAWs were used in all calculations with valence electrons being considered for C atoms and termination groups (O, F, S, Cl). For Zr atoms, the semi-core $s$ and $p$ states were added ($4s^2 4p^6 4d^2 5s^2$). For optimizations of the unit cell, k-point grid density $12\times 12\times 1$ was used and a denser $18\times 18\times 1$ grid was used for subsequent ground-state calculations. In all band structure calculations, the standard k-point path of $\Gamma$-M-K-$\Gamma$ for 2D hexagonal systems was used.

\textit{Ab initio} molecular dynamics (AIMD) was performed on $6\times 6 \times 1$ supercells at 300 K, using the Andersen thermostat and the time step of 2 fs for a total of 1000 steps. Supercells were generated by the \texttt{Phonopy} code.\cite{Togo2015}

The quasi-particle energies $\varepsilon_{n\bm{k}}^{\mathrm{GW}}$ were computed as the first-order corrections to the Kohn-Sham energies $\varepsilon_{n\bm{k}}$ (so called G$_0$W$_0$ variant).\cite{Shishkin2006} The quasi-particle gap was then calculated as $\Delta^\mathrm{GW}=\epsilon_\mathrm{CBM}^{\mathrm{GW}}-\epsilon_\mathrm{VBM}^{\mathrm{GW}}$, where VBM denotes valence band maximum, and CBM stands for conduction band minimum. It has been shown, that this GW approach can for 2D materials produce very accurate results (when the method is well-converged).\cite{Kolos2019,Dubecky2020,Kolos2022} The BSE was then used in an eigenvalue problem form\cite{Albrecht1998} for insulating materials with occupied valence bands ($v$ index), and completely unoccupied conduction bands ($c$ index),
\begin{equation}
\label{BSE}
\begin{multlined}
(\epsilon_{c\bm{k}}^{\mathrm{GW}}-\epsilon_{v\bm{k}}^{\mathrm{GW}})A_{cv \bm{k}}^\lambda
+\sum_{c'v'\bm{k}'}[2\langle \phi_{c\bm{k}}\phi_{v\bm{k}} \lvert \nu \rvert \phi_{c'\bm{k}'}\phi_{v'\bm{k}'}\rangle \\ 
-\langle \phi_{c\bm{k}}\phi_{c'\bm{k}'} \lvert W \rvert \phi_{v\bm{k}}\phi_{v'\bm{k}'}\rangle
]A_{c'v'\bm{k'}}^\lambda
=
E_{\mathrm{exc}}^\lambda A_{cv\bm{k}}^\lambda,
\end{multlined}
\end{equation}
where $\nu$ is the Coulomb kernel, $1/\lvert \bm{r}-\bm{r}'\rvert$, $W$ is the dynamically screened potential, and the eigenvectors $A_{cv \bm{k}}^\lambda$ correspond to the amplitudes of free electron-hole pair configurations composed of electron states $\rvert \phi_{c\bm{k}}\rangle$ and hole states $\rvert \phi_{v\bm{k}}\rangle$. The eigenenergies $E_{\mathrm{exc}}^\lambda$ correspond to the excitation energies (with optical gap $\Delta_\mathrm{opt}^\mathrm{BSE} \equiv E_{\mathrm{exc}}^{\lambda}$, for $\lambda$ from first nonzero transition, i.e., first bright exciton). 
The difference $E_\mathrm{b}=\Delta^{\mathrm{GW,dir}}-\Delta^{\mathrm{BSE}}_{\mathrm{opt}}$ is called exciton binding energy, where "dir" index denotes direct quasiparticle gap.

During the convergence of GW parameters, we observed that the quasiparticle gap and the optical gap converged slowly with increasing height of the computational cell $L_z$. To ensure more credible results, we fitted the values with the linear fit\cite{Karlicky2018, Dubecky2023, Berseneva2013, Choi2015}
\begin{equation}
    \Delta_{\mathrm{f}}^{\mathrm{GW}}\left(\frac{1}{L_z}\right) = C\frac{1}{L_z} + \Delta_{\mathrm{f}}^{\mathrm{GW}}(0),
\end{equation}
where $C$ and $\Delta_{\mathrm{f}}^{\mathrm{GW}}(0)$ are fitting parameters. The extrapolation to the zero $1/L_z$ limit was then used to form a corresponding \textit{a posteriori} rigid correction by taking $\Delta_{\mathrm{f}}^{\mathrm{GW}}(0)-\Delta_{\mathrm{f}}^{\mathrm{GW}}(1/L_z)$, where $1/L_z=0.05 \ \ag^{-1}$. Similarly, we fitted the values of the optical gap $\Delta_{\mathrm{opt}}^{\mathrm{BSE}}$ and formed \textit{a posteriori} correction of $\Delta_{\mathrm{opt}}^{\mathrm{BSE}}(0)-\Delta_{\mathrm{opt}}^{\mathrm{BSE}}(1/L_z)$ with the corresponding height of the computational cell. Therefore, all values of the quasiparticle gap and optical gap reported in the following sections are corrected. The results of convergence calculations for the direct and indirect quasiparticle band gaps and optical gaps depending on the height of the computational cell $L_z$ can be found in Figure S1 in the \textcolor{blue!70!black}{supplementary material}. Other GW parameters, for which the values converged sufficiently, were set as follows -- the total number of bands $N_{\mathrm{B}}=768$, the energy cutoff for the response function $E_{\mathrm{cut}}^{\mathrm{GW}}=200$ eV, the number of frequency-dependent grid points $N_{\omega}=192$ and the $18 \times 18 \times 1$ k-point grid consistent with the rest of the ground-state calculations.

\section{\label{sec:res}Results and Discussion}

\subsection{ENERGETICS AND STABILITY}

\begin{figure}[ht]
\centering
  \includegraphics[height=11cm]{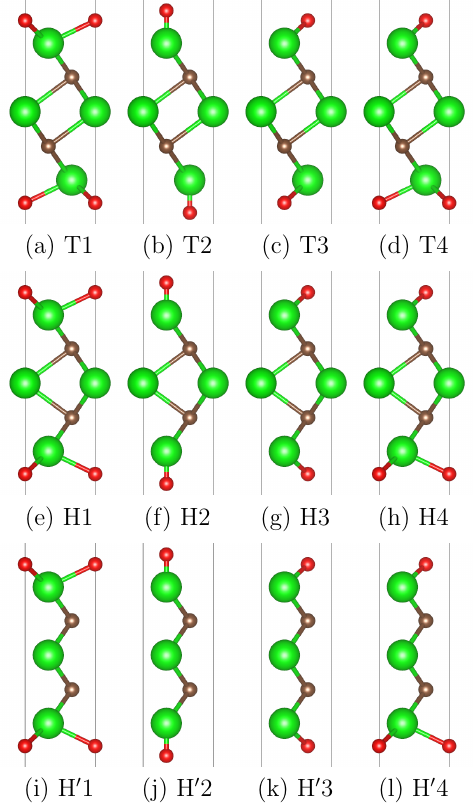}
  \caption{T-, H-, and H$^\prime$- geometric conformations of \ce{Zr3C2T2} with varying positions (1-4) of termination groups. Green atoms = Zr, brown = C, red = O, F, S or Cl. Vertical lines depict the boundaries of the unit cell. Two-layered MXenes \ce{Zr2CT2} then possess the T- and H- geometries accordingly.}
  \label{geom_isomers}
\end{figure}

We considered 8 different high-symmetric configurations for the two-layered \ce{Zr2CT2} MXene, and 12 geometries for three-layered \ce{Zr3C2T2} MXene (see Figure \ref{geom_isomers}, only the three-layered case is shown).
The conformations for the two-layered \ce{Zr2CT2} MXene correspond to the T- (trigonal) and H- (hexagonal) structures of transition metal dichalcogenides.\cite{Qian2014} For three-layered \ce{Zr3C2T2} MXene we consider one trigonal (T) structure of space group $D_{3d}$ and two hexagonal (H, H$^\prime$) structures of space group $D_{3h}$. 
The numbers in Figure \ref{geom_isomers} then denote the positions of functional groups (O, F, S, Cl): hollow site (1), metal site (2), carbon site (3), and mixed site (4). 
After optimizing the structures at the level of PBE and SCAN density functionals we conclude, that every MXene, regardless of the layers, has a trigonal T1 ground state. 
The results for the SCAN functional are collected in Table \ref{SCAN-results}. The O- and S- terminated Zr-MXenes have a distinct ground state with the second most favorable conformation being $>0.5$ eV higher in total energy (similarly as its Ti- and Hf-based analogs\cite{Kumar2023, Kumar2024}). This energy difference ensures that the ground-state conformation stays energetically more favorable even at a non-zero temperature and that no switching of phases occurs. For F- and Cl- terminated MXenes, on the other hand, two different conformations must be taken into account as the energy differences between the ground-state and the second most favorable conformations are much smaller ($0.002 - 0.06$ eV, see Table \ref{SCAN-results}). Therefore, for F- and Cl- terminated MXenes, T1 and T4 configurations are chosen for the two-layered cases, while T1 and H1 configurations are chosen for the three-layered MXenes. 

\begin{table}[ht]
    \centering
    \caption{Relative energies (in eV) of two-layered (Zr$_2$CT$_2$) and three-layered (Zr$_3$C$_2$T$_2$) Zr-based MXenes in various conformations and with changing termination groups (T = O, F, S, Cl) at the level of SCAN density functional.} 
    \def\arraystretch{1.3}%
    \begin{tabular}{cccccc}%
    \hlineB{3}
        ~ & ~ & O & F & S & Cl \\ \hlineB{3}
        \multirow{8}{3em}{\makecell{Zr$_2$C}} & T1 & 0.00 & 0.00 & 0.00 & 0.00 \\
        ~ & T2 & -$^a$ & -$^a$ & -$^a$ & -$^a$ \\ 
        ~ & T3 & 2.43 & 0.44 & 1.21 & 0.34 \\ 
        ~ & T4 & 1.03 & 0.06 & 0.51 & 0.06 \\ 
        ~ & H1 & 0.70 & 1.45 & 1.18 & 1.76 \\ 
        ~ & H2 & 8.17 & 4.68 & 6.49 & -$^b$ \\ 
        ~ & H3 & 2.95 & 1.99 & 2.23 & 2.06 \\ 
        ~ & H4 & 1.69 & 1.67 & 1.65 & 1.89 \\ \hline
        \multirow{12}{3em}{\makecell{Zr$_3$C$_2$}} & T1 & 0.00 & 0.00 & 0.00 & 0.00 \\
        ~ & T2 & 7.55 & 2.35 & 4.72 & 2.44 \\
        ~ & T3 & 1.91 & 0.71 & 0.94 & 0.45 \\
        ~ & T4 & 0.95 & 0.33 & 0.49 & 0.21 \\
        ~ & H1 & 0.53 & 0.002 & 0.54 & 0.03 \\
        ~ & H2 & 8.11 & -$^b$ & 4.89 & 2.50 \\
        ~ & H3 & 2.91 & 0.91 & 1.68 & 0.62 \\
        ~ & H4 & 1.70 & 0.44 & 1.11 & 0.32 \\
        ~ & H$^\prime$1 & 2.34 & 2.66 & 2.98 & 3.01 \\
        ~ & H$^\prime$2 & 24.26 & -$^b$ & -$^b$ & 5.77 \\
        ~ & H$^\prime$3 & 4.36 & 3.55 & 3.92 & 3.46 \\
        ~ & H$^\prime$4 & 3.29 & 3.08 & 3.43 & 3.21 \\ \hlineB{3}
    \end{tabular}
    \begin{flushleft}
    {\footnotesize $^a$These systems converged to a more stable T1 configuration during the optimization process.}\par
    {\footnotesize $^b$Conformers have proven to be unstable at the optimization step.}
    \end{flushleft}
    \label{SCAN-results}
\end{table}

To determine the stability of given systems, and to possibly resolve the energetically close isomers, we have conducted \textit{ab initio} molecular dynamics (AIMD) simulations at 300 K on selected energetically the most favorable Zr-based MXenes, as discussed above. The resulting AIMD energy profiles can be seen in Figure S2 in the \textcolor{blue!70!black}{supplementary material}. No Zr-based MXene transformed into a different conformation or exhibited signs of structural instability that would disrupt the material.

\subsection{ELECTRONIC PROPERTIES AND STRAIN}
We performed calculations on Zr-based MXenes with the inclusion of the spin-orbit coupling (SOC), which are important in analogous Hf-based MXenes.\cite{Kumar2024} The results, however, show that the SOC has only a marginal effect on the total energies and band structures of these MXenes (see Figure S3 with the \ce{Zr2CO2} MXene in the \textcolor{blue!70!black}{supplementary material}). To lower the computational cost we performed subsequent calculations without the inclusion of SOC.

\begin{figure}[ht]
\centering
\includegraphics[width=8.5cm]{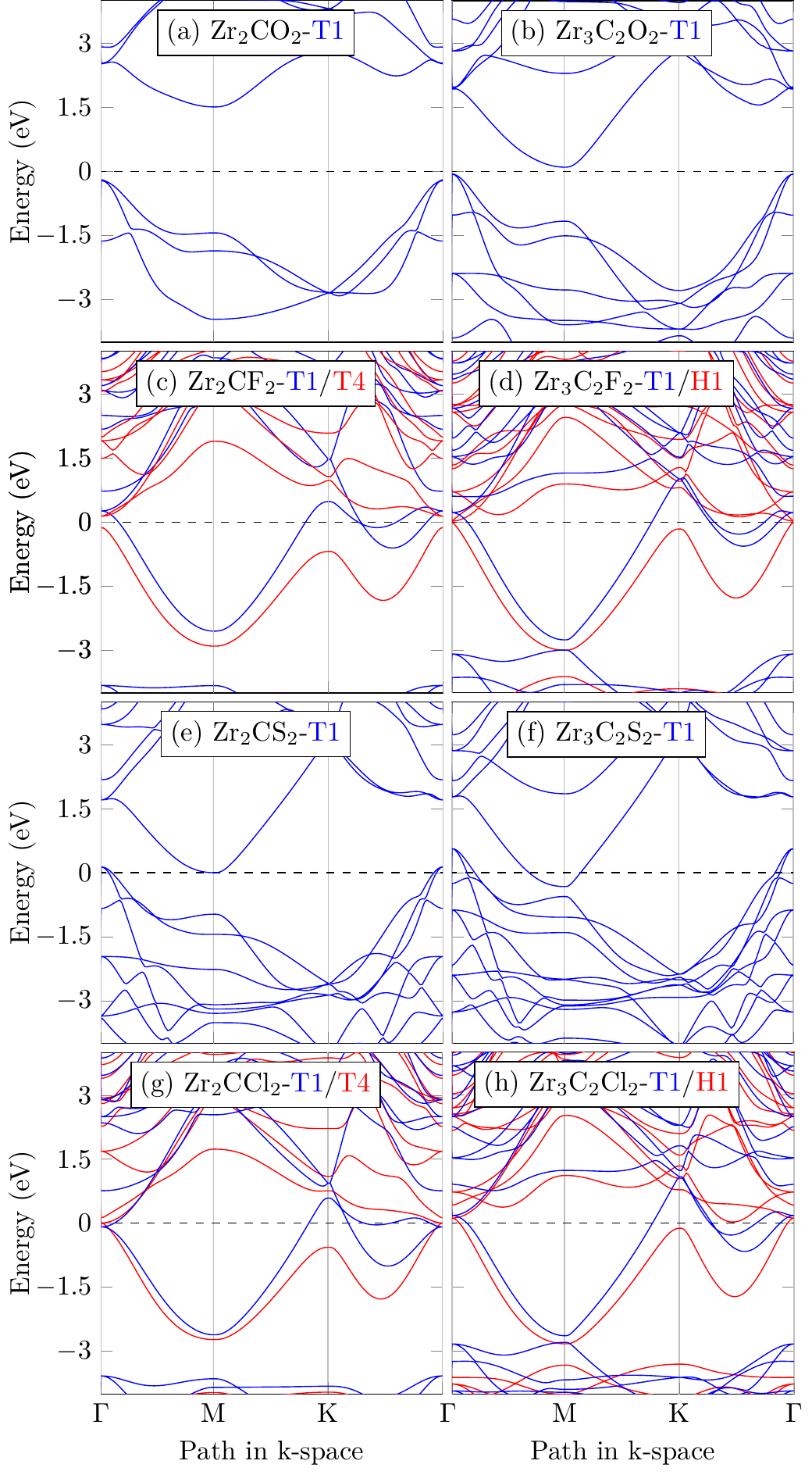}
\caption{HSE06 band structures for Zr-based MXenes Zr$_2$CT$_2$ and Zr$_3$C$_2$T$_2$, where T = O, F, S, Cl. Two band structures for different geometric conformations are presented when the given conformations are energetically very close (<0.07 eV).}
\label{band-structures}
\end{figure}

Upon analyzing the most favorable conformations, considering two possible structures for F- and Cl- terminated MXenes, some of the SCAN band structures raised a suspicion that the meta-GGA functional might be insufficient for the accurate determination of the band structures and for the evaluation of electronic band gaps in semiconducting MXenes. Therefore a set of HSE06 band structures for chosen MXenes is presented in Figure \ref{band-structures}. SCAN band structures can be seen in Figure S4 in the \textcolor{blue!70!black}{supplementary material}, as well as the comparison of meta-GGA and hybrid band structures in Figure S5 of the \textcolor{blue!70!black}{supplementary material}. At the level of hybrid density functional the number of semiconducting cases has risen significantly when compared to the available literature, where only \ce{Zr2CO2} is usually identified as a semiconductor. Here we show, that when conformations energetically close to the ground state are considered, we can identify a total of 5 semiconducting Zr-based MXenes - \ce{Zr2CO2}-T1, \ce{Zr2CF2}-T4, \ce{Zr2CCl2}-T4, \ce{Zr3C2O2}-T1, and \ce{Zr3C2F2}-H1. Note that the semiconducting conformations for F- and Cl-terminated MXenes are in Figure \ref{band-structures}c, \ref{band-structures}d, and \ref{band-structures}g shown in red. The band gaps for selected semiconducting MXenes obtained with different levels of computational theory (PBE, SCAN, HSE06, and G$_0$W$_0$@HSE06) are collected in Table \ref{band-gaps}.

\begin{table}[ht]
    \centering
    \caption{Band gaps of semiconducting Zr-based MXenes at different levels of computational theory (in eV). For indirect materials a value of the direct band gap is given in parentheses.}
    \def\arraystretch{1.6}%
    \begin{tabular}{p{0.3cm}p{0.1cm}p{0.1cm}p{0.8cm}p{0.8cm}p{0.8cm}p{0.8cm}}%
    \hlineB{3}
        ~ & ~ & ~ & PBE & SCAN & HSE06 & G$_0$W$_0^a$ \\ \hlineB{3}
        \multirow{3}{1em}{\makecell{Zr$_2$C}} & O & T1 & 0.95(1.81) & 1.25(2.11) & 1.73(2.72) & 2.61(3.56) \\ 
        ~ & F & T4 & 0.22 & 0.26 & 0.26 & 0.72 \\ 
        ~ & Cl & T4 & $0.00^b$ & $0.00^b$ & $0.01$ & 0.25$^c$ \\ \hline 
        \multirow{2}{1em}{\makecell{Zr$_3$C$_2$}} & O & T1 & - & - & 0.17(1.27) & 0.61(1.71) \\ 
        ~ & F & H1 & - & 0.05 & 0.03 & 0.29 \\ 
        \hlineB{3}
    \end{tabular}
    \begin{flushleft}
    {\footnotesize $^a$For the G$_0$W$_0$, HSE06 wave functions were used, and the reported value assumes \textit{a posteriori} rigid correction due to the height of the computational cell $L_z$.}\par
    {\footnotesize $^b$The zero reported means that the system is a zero-gap semiconductor.}\par
    {\footnotesize $^c$The rigid correction was fine-tuned on +2\% strained system.}
    \end{flushleft}
    \label{band-gaps}
\end{table}

\begin{figure}[ht]
\centering
\includegraphics[width=8.6cm]{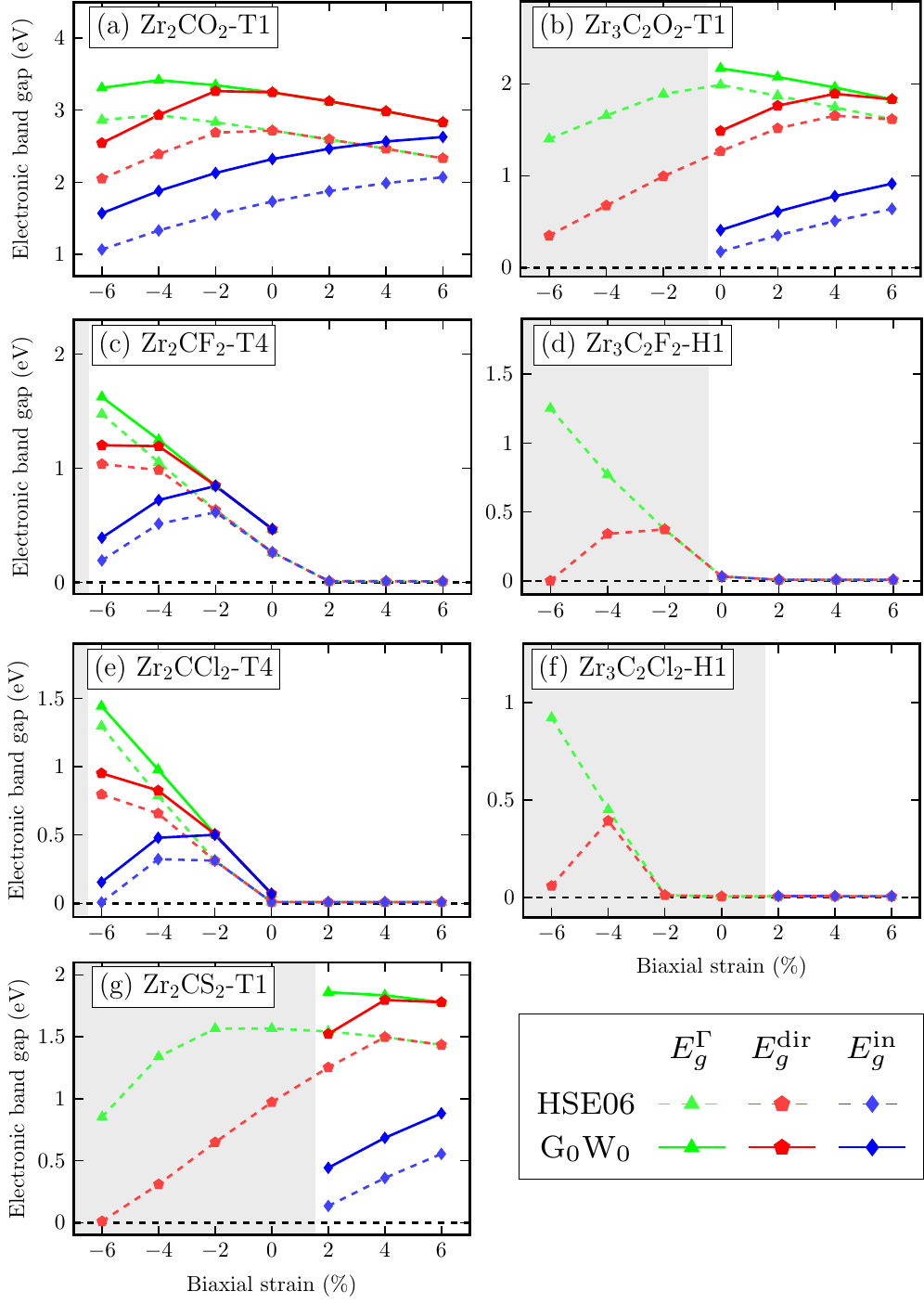}
\caption{The effect of biaxial strain on the indirect ($E_{\mathrm{g}}^{\mathrm{in}}$), direct ($E_{\mathrm{g}}^{\mathrm{dir}}$) and $\Gamma$-point ($E_{\mathrm{g}}^{\mathrm{\Gamma}}$) electronic band gap in several Zr-based MXenes. The MXenes, which exhibited conducting behavior in the full strain range, are not shown. For chosen MXenes, two levels of theory are used to model the behavior under biaxial strain (hybrid functional HSE06 and many-body perturbation G$_0$W$_0$@HSE06 method). The grey area signifies the biaxial strain beyond which the material lost its indirect band gap and became conducting at the hybrid density functional level.}
\label{strain-results}
\end{figure}

The changes in indirect, direct, and $\Gamma$-point electronic band gap for chosen MXenes under biaxial strain are shown in Figure \ref{strain-results}. MXenes that are missing from this selection exhibited conducting behavior in the full strain range. Each MXene presented in Figure \ref{strain-results} has been tested in AIMD simulation in the range from -6\% to +6\% biaxial strain to ensure their stability in these regions. Individual band structures and their changes with biaxial strain for every configuration can be seen in Figures S6 -- S17 in the \textcolor{blue!70!black}{supplementary material}. The changes in the total energy of given conformations under strain are presented in Figure S18 of the \textcolor{blue!70!black}{supplementary material}. The detailed analysis of the electronic properties of given Zr-based MXenes and their changes with the biaxial strain are presented in the following sections, with a focus on each termination group separately.

\begin{figure}[ht]
\centering
\includegraphics[width=8.5cm]{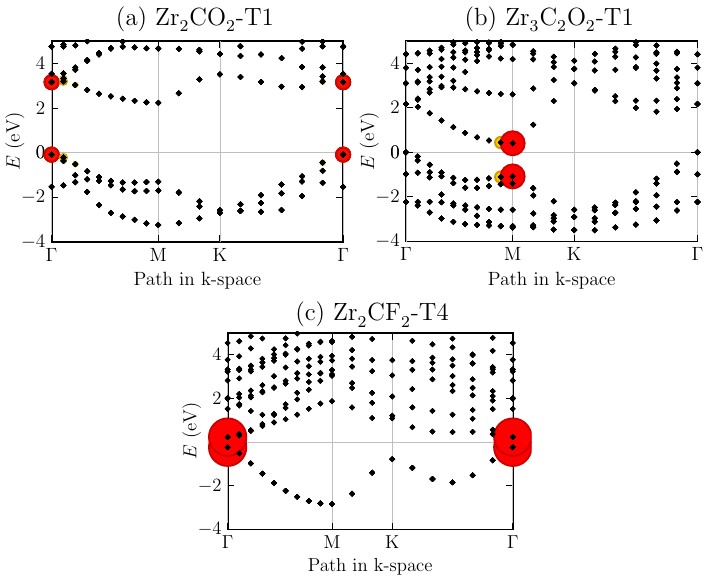}
\caption{Quasiparticle (GW) band structure (black dots) of (a) \ce{Zr2CO2}-T1, (b) \ce{Zr3C2O2}-T1, and (c) \ce{Zr2CF2}-T4 and all $|A_{cv \bm{k}}^\lambda|$ coefficients from BSE (represented by the radii of the colored circles) visually show which electron-hole pairs contribute to the first bright excitonic peak. The top of the valence band is set to zero.}
\label{fatbands}
\end{figure}

\subsubsection{O-terminated Zr-based MXenes}
In an unstrained state the O-terminated two-layered MXene \ce{Zr2CO2} exhibits at levels of PBE, SCAN, HSE06, and G$_0$W$_0$@HSE06 an indirect band gap 0.95 eV, 1.25 eV, 1.73 eV, and 2.61 eV, respectively, and a direct band gap 1.81 eV, 2.11 eV, 2.72 eV, and 3.56 eV, respectively (see Table \ref{band-gaps}). 
The indirect gap is located between $\Gamma$ and M points in the Brillouin zone, while the minimal direct gap is located in the $\Gamma$ point. 
This behavior holds at all computational levels, $cf.$ the G$_0$W$_0$ band structure in Figure \ref{fatbands}a and the HSE06 band structure in Figure \ref{band-structures}a (or also the SCAN band structure in Figure S4a).  
At the level of PBE density functional, the electronic band gap of 0.88 eV is usually reported for \ce{Zr2CO2}. This result, calculated by Khazaei et al.,\cite{Khazaei2012} differs slightly from our estimation and a possible explanation could arise from the fact that four more inner electrons ($s$ and $p$ states) for Zr atoms were included in self-consistent calculations. Our results are in very good agreement with a different study done by Zha et at. where the electronic configuration for Zr atoms is identical to ours and an indirect band gap of 0.97 eV is reported for \ce{Zr2CO2}.\cite{Zha2015}

Results for \ce{Zr2CO2} indicate that a indirect-to-direct semiconductor transition would occur under tensile strain larger than +6\% (see Figure \ref{strain-results}a).
Under compressive strain, it experiences a decrease of the indirect band gap and it is expected to lose its semiconducting properties under large enough compressive strain.
These results are consistent with different studies of the effects of strain on \ce{Zr2CO2}.\cite{Khan2017,Cui2019}

With the use of SCAN density functional, the three-layered MXene \ce{Zr3C2O2} shows conducting behavior in its unstrained state (see Figure S4b in the \textcolor{blue!70!black}{supplementary material}) but it was suspected that this could be just an artifact of the GGA and meta-GGA band gap underestimation. This is subsequently proven as with the HSE06 functional the \ce{Zr3C2O2} MXene exhibits an indirect and direct electronic band gap of 0.17 eV and 1.27 eV, respectively (Figure \ref{band-structures}b). Furthermore, with the G$_0$W$_0$ the band gap increases and is estimated to be 0.61 eV (indirect between $\Gamma$ and M points) and 1.71 eV (direct in M, Figure \ref{fatbands}). Under compressive strain, the MXene becomes conductive with either density functional used. Under tensile strain, the indirect band gap increases (the material becomes semiconducting even at the SCAN level), and the direct band gap is moved to the $\Gamma$ point. 

To better understand strain-induced changes in the band structure of \ce{Zr3C2O2}-T1 MXene, we have analyzed partial charge densities of states near the band gap. These are collected in Figure S19 in the \textcolor{blue!70!black}{supplementary material}. These representations then reflect well the band structure changes in Figure S12 where the top two valence bands degenerate in the $\Gamma$ point but exchange with the increasing tensile strain, and thus we observe the change of $p$-state in the M point. Furthermore, Figure S20 in the \textcolor{blue!70!black}{supplementary material} shows the projected density of states (PDOS) of \ce{Zr3C2O2}-T1, which underline that Zr-atoms mainly contribute to the conduction states and C-atoms contribute to the valence states.

\subsubsection{F-terminated Zr-based MXenes}
The trigonal T1 geometry of the two-layered MXene \ce{Zr2CF2} is energetically more favorable in the full strain range at the meta-GGA level of the theory (see Figure S18c in the \textcolor{blue!70!black}{supplementary material}). 
On the other hand, with increasing levels of theory, energy differences between energies of conformations became even smaller: competing T1 and T4 configurations have energy differences of 0.08 eV, 0.06 eV, and 0.03 eV for GGA (PBE), meta-GGA (SCAN), and hybrid (HSE06) DFT method, respectively. Speculatively, a higher, more correlated level of theory (fully beyond the scope of our work, such as random phase approximation or quantum Monte Carlo) could, therefore, lead to the T4 ground state.
Unstrained T1 geometry is conducting and this behavior is not changed under any strain, either compressive or tensile (see Figure S7 of the \textcolor{blue!70!black}{supplementary material}). The trigonal T4 conformer exhibits at the level of the hybrid HSE06 functional a direct band gap of 0.26 eV which then gets substantially larger (0.72 eV) with the many-body G$_0$W$_0$ method (and with the correction to $L_z$). The minimal direct gap is located in the $\Gamma$ point and this behavior holds at all computational levels, $cf.$ the G$_0$W$_0$ band structure in Figure \ref{fatbands}c and the HSE06 band structure in Figure \ref{band-structures}c. Our results agree very well with another study done by Duan et al.\cite{Duan2023} where the order of conformers is equivalent to ours and the electronic band gap of 0.24 eV is reported for the \ce{Zr2CF2}-T4 MXene. 

With a compressive strain, the direct band gap of the T4 conformer becomes even larger up to -2\% strain when the MXene undergoes a transition to an indirect material (see Figure \ref{strain-results}c). From the presented results we can suspect, that beyond -6\% compressive strain the \ce{Zr2CF2}-T4 MXene would lose its band gap and become a conductor.

For the three-layered MXene \ce{Zr3C2F2}, the energy differences between T1 and H1 conformers are very small. At the ground-state SCAN level, the difference is only 0.002 eV (Table \ref{SCAN-results}). It is also the only tested MXene that experiences a phase shift under compressive strain (Figure S18d in the \textcolor{blue!70!black}{supplementary material}). This MXene, as well as \ce{Zr3C2O2}, was the first indication that a higher level density functional needs to be used. With the use of hybrid functional HSE06, the semiconducting hexagonal H1 geometry becomes the ground-state conformation. The energy levels between individual phases (T1 and H1) then get pushed further apart as the H1 geometry is 0.02 eV energetically lower than the T1 geometry.

The T1 geometry exhibits a metallic behavior in the full range of compressive and tensile strain regardless of the method used (Figure S13 in the \textcolor{blue!70!black}{supplementary material}). The hexagonal H1 geometry has at its unstrained state a small direct band gap of 0.03 at the hybrid level. Under a relatively small -2\% compressive strain, the \ce{Zr3C2F2}-H1 MXene starts exhibiting conducting behavior (see Figure \ref{strain-results}d). Under tensile strain, the band gap vanishes to zero (similarly as in the two-layered case) and the H1 conformer stays as a zero-gap semiconductor further on. These near-zero band gaps (which occur under tensile strain in all tested F- and Cl-terminated MXenes, see Figure \ref{strain-results}c-f) have proven to be quite challenging when one chooses to use the many-body approach. With valence and conduction bands being so close to the Fermi level, the G$_0$W$_0$ method can sometimes make a wrong correction and predict that material, claimed as semiconducting at the SCAN or HSE06 level, is conducting. One thus has to be careful in such scenarios, as the wrong initial guess, such as slightly partially occupied bands near the Fermi level, can get magnified and result in the wrong final prediction. In the case of this MXene, the SCAN band gap of 0.05 eV was with the use of G$_0$W$_0$@SCAN wrongfully predicted as conductor. The HSE06 band gap of 0.03 eV was then successfully corrected to 0.29 eV after the G$_0$W$_0$@HSE06 computation (taking into account the correction to $L_z$).

\subsubsection{S-terminated Zr-based MXenes}
The two-layered MXene \ce{Zr2CS2} exhibits conducting behavior while unstrained. However, the emergence of a non-zero gap can be observed under tensile strain beyond +2\% at the hybrid functional level (see Figure \ref{strain-results}g). The G$_0$W$_0$@HSE06 approach then suggests that \ce{Zr2CS2} could retain its semiconducting properties even in its unstrained state. However, without the initial semiconducting guess at no strain, we are not able to confirm this hypothesis. Under compressive strain, the MXene retains its conducting properties.\\

Three-layered \ce{Zr3C2S2} MXene exhibits conductive behavior in the full strain range and doesn't show signs, that a band gap could appear (Figure S15 in the \textcolor{blue!70!black}{supplementary material}).

\subsubsection{Cl-terminated Zr-based MXenes}
T1 geometry of the two-layered MXene \ce{Zr2CCl2} remains energetically more favorable (0.03 eV or less) for the full range of compressive and tensile strain (see Figure S18g in the \textcolor{blue!70!black}{supplementary material}). The hybrid HSE06 density functional then pushes the energy levels closer but the order remains unchanged (and we can speculate on almost identical energies using a higher, more correlated, level of theory). The same holds for the three-layered Cl-terminated MXene \ce{Zr3C2Cl2}.

Unstrained \ce{Zr2CCl2}-T1 conformer exhibits conducting behavior and this behavior is unchanged under any strain (Figure S10 in the \textcolor{blue!70!black}{supplementary material}). The T4 geometry has at the level of hybrid density functional a very small direct band gap of 0.01 eV. This band gap is then increased by the many-body G$_0$W$_0$ approach to 0.25 eV (after rigid correction to $L_z$). Under tensile strain, the direct band gap diminishes to near-zero values, as in other halogen-terminated MXenes reported here (Figure \ref{strain-results}e). Under compressive strain, the MXene behaves similarly to the F-terminated \ce{Zr2CF2}-T4 (see Figure \ref{strain-results}c and \ref{strain-results}e). The direct band gap increases up to -2\% strain, the material then becomes indirect, and with further strain, the band gap disappears as the valence band crosses the Fermi level.

The three-layered T1 geometry of the \ce{Zr3C2Cl2} MXene is metallic in its unstrained state and the behavior is not changed under either compressive or tensile strain (Figure S16 in the \textcolor{blue!70!black}{supplementary material}). The H1 geometry is also predicted to be conducting in its unstrained state (Figure \ref{strain-results}f). Under a small +2\% tensile strain, we can observe the disappearance of the conducting behavior as there again appears to be a near-zero band gap at the $\Gamma$ point. Under compressive strain, the \ce{Zr3C2Cl2} MXene retains its conducting properties.

\subsection{OPTICAL PROPERTIES}
\begin{figure}[ht]
    \centering
    \includegraphics[width=8.5cm]{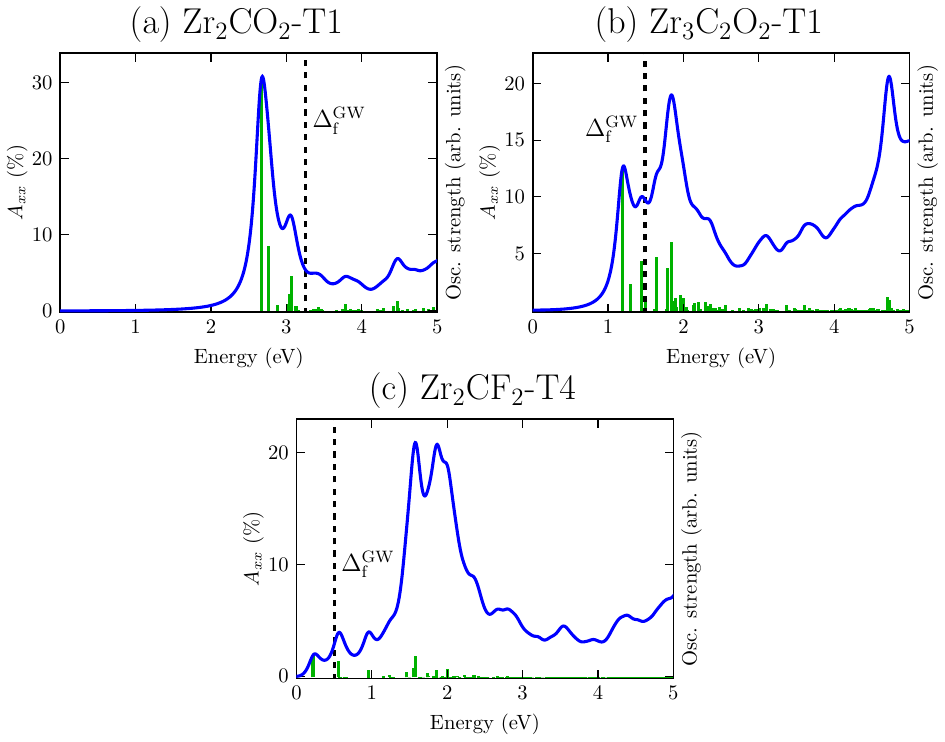}
    \caption{Optical absorption spectra ($A_{xx}=A_{yy}$, $A_{zz}\approx 0$) for Zr-based MXenes (a) \ce{Zr2CO2}-T1, (b) \ce{Zr3C2O2}-T1, and (c) \ce{Zr2CF2}-T4 at the level of GW@HSE06+BSE and corresponding oscillator strengths. The dashed line in spectra marks the GW+BSE estimation of the fundamental band gap $\Delta_{\mathrm{f}}$.}
    \label{spectra}
\end{figure}

We provide calculated absorption spectra for three selected Zr-based MXenes, \ce{Zr2CO2}-T1, \ce{Zr3C2O2}-T1, and \ce{Zr2CF2}-T4, by the GW+BSE many-body approach. The very small band gap of the remaining semiconducting MXenes made the estimation of the optical gap unreliable. The number of occupied and virtual bands that enter the solution of BSE was chosen to ensure credibility up to a photon energy of 5 eV. The resulting value of optical gap $\Delta_{\mathrm{opt}}^{\mathrm{BSE}}$ for each MXene is also, similarly as for quasiparticle gap $\Delta_{\mathrm{f}}^{\mathrm{GW}}$, corrected by \textit{a posteriori} rigid correction, as discussed in Section \ref{sec:methods}. The resulting absorption spectra, paired with corresponding oscillator strengths, can be seen in Figure \ref{spectra} with the quasiparticle band gap marked with a dashed line. The visualized quantity here is the optical absorptance $A(E) = 1-\mathrm{exp}[-\epsilon_2 E L_z/{\hbar}c]$,\cite{Ketolainen2020} where $E$ is the energy of incident photon, $\epsilon_2$ is the imaginary part of dielectric function, $\hbar$ is reduced Planck's constant, and $c$ is speed of light.

\begin{table}[h]
    \centering
    \caption{The direct quasiparticle band gap ($\Delta_{\mathrm{f}}^{\mathrm{GW,dir}}$), optical gap ($\Delta_{\mathrm{opt}}^{\mathrm{BSE}}$), and the exciton binding energy ($E_{\mathrm{b}}$) for three selected Zr-based MXenes.}
    \def\arraystretch{1.3}%
    \begin{tabular}{cccc}%
    \hlineB{3}
        ~ & $\Delta_{\mathrm{f}}^{\mathrm{GW,dir}}$ & $\Delta_{\mathrm{opt}}^{\mathrm{BSE}}$ & $E_{\mathrm{b}}$ \\ \hlineB{3}
        \ce{Zr2CO2}-T1 & 3.56 & 2.56 & 1.00 \\ 
        \ce{Zr3C2O2}-T1 & 1.71 & 1.09 & 0.62 \\ 
        \ce{Zr2CF2}-T4 & 0.72 & 0.20 & 0.52   \\ \hlineB{3}
    \end{tabular}
    \label{opt_table}
\end{table}

For all selected MXenes, we can observe at least partial absorption of the visible range of incoming radiation (1.63-3.26 eV) with exceptionally high efficiency in a range of 20-30\%. Table \ref{opt_table} collects the many-body results for the three selected MXenes, including the direct quasiparticle gap, optical gap, and their difference in the form of exciton binding energy. 
The first bright exciton (i.e., first nonzero optical transition) occurs for $\lambda = $ 2, 4, and 1 in \ce{Zr2CO2}-T1, \ce{Zr3C2O2}-T1, and \ce{Zr2CF2}-T4, respectively. 
The dominant contribution to the corresponding excitonic wave function (expressed in an electron-hole product basis as $\sum_{cv \bm{k}} A_{cv \bm{k}}^\lambda \phi_{c\bm{k}}\phi_{v\bm{k}}$; see also equation \ref{BSE}) is always from k-space area sharply located in minimal direct gap position (i.e., $\Gamma$, M, and $\Gamma$ location for \ce{Zr2CO2}-T1, \ce{Zr3C2O2}-T1, and \ce{Zr2CF2}-T4, respectively). 
This is visible in Figure \ref{fatbands} by color circles representing $|A_{cv \bm{k}}^\lambda|$ as the important electron-hole contributions labeled by $c,v$ subscripts (band pairs on the y-axis; only the valence band and conduction band are important) and by $\bm{k}$ subscript describing the x-axis (missing color circles on black points means negligible coefficients are not visible in the graph). 

It is possible to compare materials based on the relationship between $\Delta_{\mathrm{f}}^{\mathrm{GW}}$ and $E_{\mathrm{B}}$ as many 2D semiconductors exhibit linear scaling $E_{\mathrm{B}}\approx\Delta_{\mathrm{f}}/4$.\cite{Jiang2017} Both O-terminated MXenes satisfy this scaling within some margin of error. The F-terminated MXene, on the other hand, exceeds this ratio three-fold. 

\begin{figure}[hbt]
    \centering
    \includegraphics[width=8.5cm]{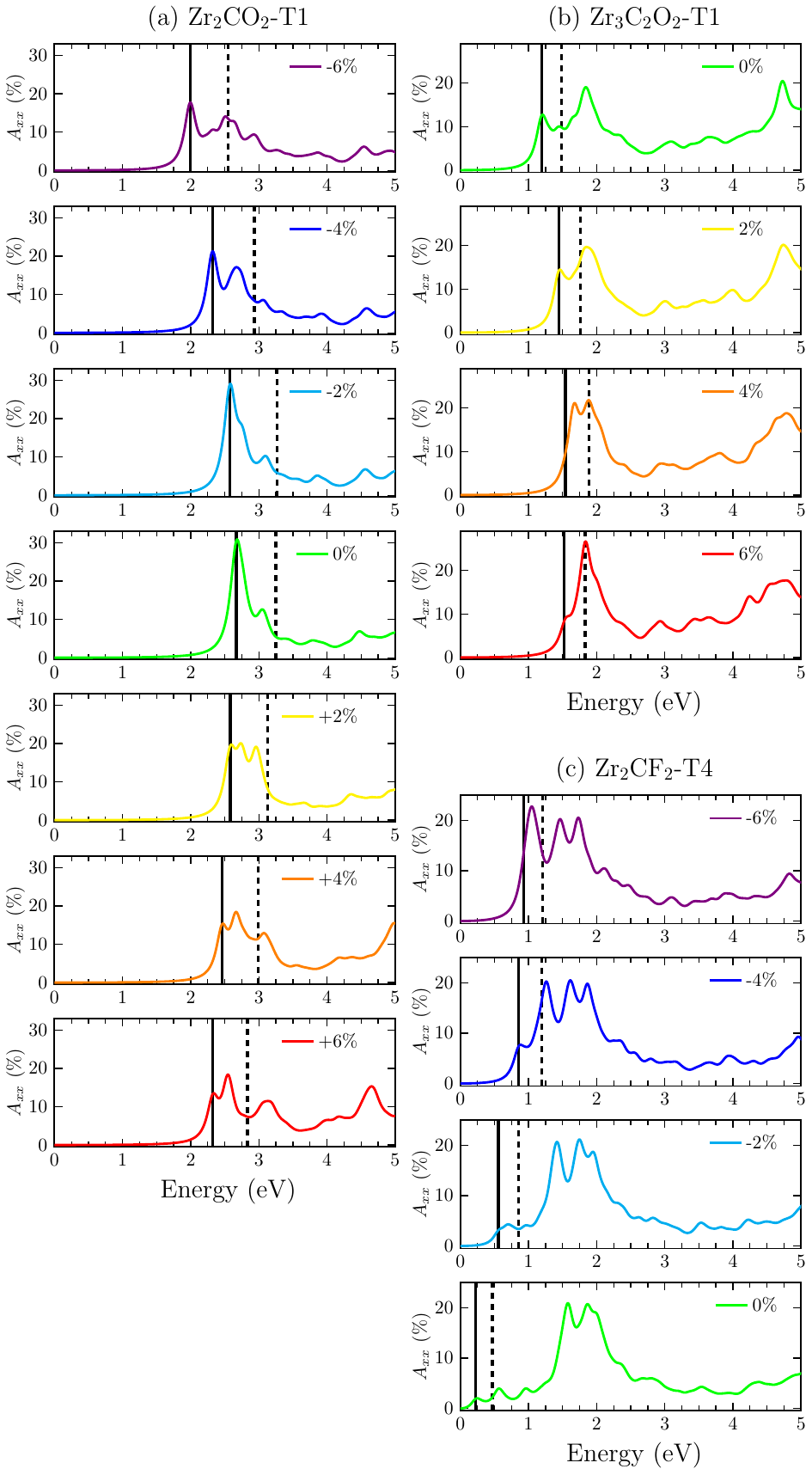}
    \caption{Optical absorption spectra ($A_{xx}=A_{yy}$, $A_{zz}\approx 0$) for Zr-based MXenes (a) \ce{Zr2CO2}-T1, (b) \ce{Zr3C2O2}-T1 and (c) \ce{Zr2CF2}-T4 computed with the GW+BSE method depended on biaxial strain. The black solid and dashed lines note the position of the first optically active exciton $\Delta_{\mathrm{opt}}^{\mathrm{BSE}}$ and the quasiparticle band gap $\Delta_{\mathrm{f}}^{\mathrm{GW}}$, respectively.}
    \label{strain_spectra}
\end{figure}

To investigate the effect of strain on the optical properties of Zr-based MXenes, we have calculated the optical absorption spectra of the three selected MXenes depending on the biaxial strain. The resulting spectra can be seen in Figure \ref{strain_spectra} with the quasiparticle gap and optical gap marked. It can be seen that the optical gap follows a pattern similar to the results collected in Figure \ref{strain-results}. Therefore no significant change to the exciton binding energy is observed. 
On the other hand, it is possible to tune the energy/wavelength region of maximal absorption by strain (Figure \ref{strain_spectra}). The \ce{Zr2CO2}-T1 MXene has the highest absorption efficiency in its unstrained state. 
When tensile or compressive strain is applied, the absorption maxima decreases but remains in the range of 15-20\%. The three-layered \ce{Zr3C2O2}-T1 MXene, on the other hand, shows an increase in absorption when tensile strain is applied. For F-terminated \ce{Zr2CF2}-T4, we observe that when the material is compressed, the first optically active exciton is shifted to higher energies (correspondingly with the change in the fundamental band gap). The bulk of absorption is at the same time shifted to lower energies.

\section{Conclusions}
In this study, we have systematically investigated the strain-induced electronic and optical properties of zirconium-based MXenes, specifically focusing on two- and three-layered configurations of \ce{Zr2CT2} and \ce{Zr3C2T2} MXenes, where T is a terminal group. Through extensive computational analyses employing methodologies of density functional theory (DFT) ranging from the Generalized Gradient Approximation (GGA), meta-GGA, to the more sophisticated hybrid functional and many-body single-shot G$_0$W$_0$ method, we have explored a total of 8 highly symmetrical configurations for two-layered and 12 configurations for three-layered Zr-based MXenes, encompassing a variety of terminations including oxygen, fluorine, sulfur, and chlorine.

Our findings reveal that all investigated Zr-based MXenes exhibit, at the meta-GGA level, a trigonal ground state with a second geometry being energetically very close to the ground state in halogen-terminated MXenes. The energy levels of different phases become more closely spaced upon using the hybrid functional, and \ce{Zr3C2F2} MXene has hexagonal semiconducting ground state. Diverging from the prevailing focus on oxygen-terminated MXenes in the available literature, which predominantly identifies only \ce{Zr2CO2} as semiconducting, our research, particularly at the level of the hybrid functional HSE06, identifies three Zr-based MXenes that are semiconducting in their ground-state conformation: \ce{Zr2CO2}, \ce{Zr3C2O2}, \ce{Zr3C2F2}. Moreover, two more MXenes, \ce{Zr2CF2}, and \ce{Zr2CCl2}, are found to have a semiconducting conformation energetically very close to the ground state, within a margin of less than 0.03 eV at hybrid density functional level (or speculatively the true ground state if using a higher, more correlated, level of theory).

Furthermore, our investigation reveals that additional conducting MXenes (\ce{Zr2CS2}, \ce{Zr3C2Cl2}) transition to a semiconducting state under a relatively minor application of tensile biaxial strain, thus expanding the potential catalog of Zr-based MXenes suitable for semiconducting applications. This finding suggests that Zr-based MXenes hold a broader utility in the fabrication of electronic devices than previously recognized. On the other hand, the semiconducting Zr-based MXenes always switch to conducting states under compressive biaxial strain.

Advancing beyond electronic properties, our study also employs the solution of the Bethe-Salpeter equation (BSE) to elucidate the optical absorption spectra, estimate the optical gap, and determine the exciton binding energy for three selected MXenes: \ce{Zr2CO2}, \ce{Zr2CF2}, and \ce{Zr3C2O2}. The results from this segment of our research highlight the exceptional potential of these MXenes to absorb electromagnetic radiation within the visible photon energy range of 1.63 - 3.26 eV, demonstrating an unusually high efficiency of 20-30\%.

The thermal stability, the possibility of semiconductor-to-conductor transitions, and the high sensitivity of Zr-based MXenes to external conditions and functionalization are unusually promising combinations. These make the Zr-based MXenes materials promising for device applications at operational temperatures in electronic, sensor, and optical technologies.

\section{Supplementary Material}
See the \textcolor{blue!70!black}{supplementary material} for computational details, more details on the geometrical structure and stability, the convergence of electronic and optical properties, and properties under strain.

\section{Acknowledgment}
This article has been produced with the financial support of the Czech Science Foundation (21-28709S), the University of Ostrava (SGS04/PrF/2024), and the European Union under the LERCO project (number CZ.10.03.01/00/22$\_$003/0000003) via the Operational Programme Just Transition. 
The calculations were performed at IT4Innovations National Supercomputing Center (e-INFRA CZ, ID:90140).

\section*{References}
\bibliography{refs}

\begin{thebibliography}{48}%
\makeatletter
\providecommand \@ifxundefined [1]{%
 \@ifx{#1\undefined}
}%
\providecommand \@ifnum [1]{%
 \ifnum #1\expandafter \@firstoftwo
 \else \expandafter \@secondoftwo
 \fi
}%
\providecommand \@ifx [1]{%
 \ifx #1\expandafter \@firstoftwo
 \else \expandafter \@secondoftwo
 \fi
}%
\providecommand \natexlab [1]{#1}%
\providecommand \enquote  [1]{``#1''}%
\providecommand \bibnamefont  [1]{#1}%
\providecommand \bibfnamefont [1]{#1}%
\providecommand \citenamefont [1]{#1}%
\providecommand \href@noop [0]{\@secondoftwo}%
\providecommand \href [0]{\begingroup \@sanitize@url \@href}%
\providecommand \@href[1]{\@@startlink{#1}\@@href}%
\providecommand \@@href[1]{\endgroup#1\@@endlink}%
\providecommand \@sanitize@url [0]{\catcode `\\12\catcode `\$12\catcode
  `\&12\catcode `\#12\catcode `\^12\catcode `\_12\catcode `\%12\relax}%
\providecommand \@@startlink[1]{}%
\providecommand \@@endlink[0]{}%
\providecommand \url  [0]{\begingroup\@sanitize@url \@url }%
\providecommand \@url [1]{\endgroup\@href {#1}{\urlprefix }}%
\providecommand \urlprefix  [0]{URL }%
\providecommand \Eprint [0]{\href }%
\providecommand \doibase [0]{http://dx.doi.org/}%
\providecommand \selectlanguage [0]{\@gobble}%
\providecommand \bibinfo  [0]{\@secondoftwo}%
\providecommand \bibfield  [0]{\@secondoftwo}%
\providecommand \translation [1]{[#1]}%
\providecommand \BibitemOpen [0]{}%
\providecommand \bibitemStop [0]{}%
\providecommand \bibitemNoStop [0]{.\EOS\space}%
\providecommand \EOS [0]{\spacefactor3000\relax}%
\providecommand \BibitemShut  [1]{\csname bibitem#1\endcsname}%
\let\auto@bib@innerbib\@empty
\bibitem [{\citenamefont {Anasori}, \citenamefont {Lukatskaya},\ and\
  \citenamefont {Gogotsi}(2017)}]{Anasori2017}%
  \BibitemOpen
  \bibfield  {author} {\bibinfo {author} {\bibfnamefont {B.}~\bibnamefont
  {Anasori}}, \bibinfo {author} {\bibfnamefont {M.~R.}\ \bibnamefont
  {Lukatskaya}}, \ and\ \bibinfo {author} {\bibfnamefont {Y.}~\bibnamefont
  {Gogotsi}},\ }\bibfield  {title} {\enquote {\bibinfo {title} {2d metal
  carbides and nitrides (mxenes) for energy storage},}\ }\href {\doibase
  10.1038/natrevmats.2016.98} {\bibfield  {journal} {\bibinfo  {journal} {Nat.
  Rev. Mater.}\ }\textbf {\bibinfo {volume} {2}},\ \bibinfo {pages} {16098}
  (\bibinfo {year} {2017})}\BibitemShut {NoStop}%
\bibitem [{\citenamefont {Hantanasirisakul}\ and\ \citenamefont
  {Gogotsi}(2018)}]{Hantanasirisakul2018}%
  \BibitemOpen
  \bibfield  {author} {\bibinfo {author} {\bibfnamefont {K.}~\bibnamefont
  {Hantanasirisakul}}\ and\ \bibinfo {author} {\bibfnamefont {Y.}~\bibnamefont
  {Gogotsi}},\ }\bibfield  {title} {\enquote {\bibinfo {title}
  {\textcolor{black}{Electronic and Optical Properties of 2D Transition Metal
  Carbides and Nitrides (MXenes)}},}\ }\href {\doibase
  https://doi.org/10.1002/adma.201804779} {\bibfield  {journal} {\bibinfo
  {journal} {Advanced Materials}\ }\textbf {\bibinfo {volume} {30}},\ \bibinfo
  {pages} {1804779} (\bibinfo {year} {2018})}\BibitemShut {NoStop}%
\bibitem [{\citenamefont {Gogotsi}\ and\ \citenamefont
  {Anasori}(2019)}]{Gogotsi2019}%
  \BibitemOpen
  \bibfield  {author} {\bibinfo {author} {\bibfnamefont {Y.}~\bibnamefont
  {Gogotsi}}\ and\ \bibinfo {author} {\bibfnamefont {B.}~\bibnamefont
  {Anasori}},\ }\bibfield  {title} {\enquote {\bibinfo {title}
  {\textcolor{black}{The Rise of MXenes}},}\ }\href {\doibase
  10.1021/acsnano.9b06394} {\bibfield  {journal} {\bibinfo  {journal} {ACS
  Nano}\ }\textbf {\bibinfo {volume} {13}},\ \bibinfo {pages} {8491--8494}
  (\bibinfo {year} {2019})}\BibitemShut {NoStop}%
\bibitem [{\citenamefont {Naguib}\ \emph {et~al.}(2011)\citenamefont {Naguib},
  \citenamefont {Kurtoglu}, \citenamefont {Presser}, \citenamefont {Lu},
  \citenamefont {Niu}, \citenamefont {Heon}, \citenamefont {Hultman},
  \citenamefont {Gogotsi},\ and\ \citenamefont {Barsoum}}]{Naguib2011}%
  \BibitemOpen
  \bibfield  {author} {\bibinfo {author} {\bibfnamefont {M.}~\bibnamefont
  {Naguib}}, \bibinfo {author} {\bibfnamefont {M.}~\bibnamefont {Kurtoglu}},
  \bibinfo {author} {\bibfnamefont {V.}~\bibnamefont {Presser}}, \bibinfo
  {author} {\bibfnamefont {J.}~\bibnamefont {Lu}}, \bibinfo {author}
  {\bibfnamefont {J.}~\bibnamefont {Niu}}, \bibinfo {author} {\bibfnamefont
  {M.}~\bibnamefont {Heon}}, \bibinfo {author} {\bibfnamefont {L.}~\bibnamefont
  {Hultman}}, \bibinfo {author} {\bibfnamefont {Y.}~\bibnamefont {Gogotsi}}, \
  and\ \bibinfo {author} {\bibfnamefont {M.~W.}\ \bibnamefont {Barsoum}},\
  }\bibfield  {title} {\enquote {\bibinfo {title} {Two-dimensional nanocrystals
  produced by exfoliation of ti3alc2},}\ }\href {\doibase
  https://doi.org/10.1002/adma.201102306} {\bibfield  {journal} {\bibinfo
  {journal} {Adv. Mater.}\ }\textbf {\bibinfo {volume} {23}},\ \bibinfo {pages}
  {4248--4253} (\bibinfo {year} {2011})}\BibitemShut {NoStop}%
\bibitem [{\citenamefont {Anasori}\ and\ \citenamefont
  {Gogotsi}(2020)}]{Anasori2020Book}%
  \BibitemOpen
  \bibinfo {editor} {\bibfnamefont {B.}~\bibnamefont {Anasori}}\ and\ \bibinfo
  {editor} {\bibfnamefont {Y.}~\bibnamefont {Gogotsi}},\ eds.,\ \href@noop {}
  {\emph {\bibinfo {title} {2D Metal Carbides and Nitrides (MXenes)}}}\
  (\bibinfo  {publisher} {Springer},\ \bibinfo {address} {Switzerland},\
  \bibinfo {year} {2020})\BibitemShut {NoStop}%
\bibitem [{\citenamefont {Anasori}\ and\ \citenamefont
  {Gogotsi}(2022)}]{Anasori2022}%
  \BibitemOpen
  \bibfield  {author} {\bibinfo {author} {\bibfnamefont {B.}~\bibnamefont
  {Anasori}}\ and\ \bibinfo {author} {\bibfnamefont {Y.}~\bibnamefont
  {Gogotsi}},\ }\bibfield  {title} {\enquote {\bibinfo {title}
  {\textcolor{black}{MXenes: trends, growth, and future directions}},}\ }\href
  {\doibase 10.1007/s41127-022-00053-z} {\bibfield  {journal} {\bibinfo
  {journal} {Graphene and 2D Materials}\ }\textbf {\bibinfo {volume} {7}},\
  \bibinfo {pages} {75--79} (\bibinfo {year} {2022})}\BibitemShut {NoStop}%
\bibitem [{\citenamefont {Lim}\ \emph {et~al.}(2022)\citenamefont {Lim},
  \citenamefont {Shekhirev}, \citenamefont {Wyatt}, \citenamefont {Anasori},
  \citenamefont {Gogotsi},\ and\ \citenamefont {Seh}}]{Lim2022}%
  \BibitemOpen
  \bibfield  {author} {\bibinfo {author} {\bibfnamefont {K.~R.~G.}\
  \bibnamefont {Lim}}, \bibinfo {author} {\bibfnamefont {M.}~\bibnamefont
  {Shekhirev}}, \bibinfo {author} {\bibfnamefont {B.~C.}\ \bibnamefont
  {Wyatt}}, \bibinfo {author} {\bibfnamefont {B.}~\bibnamefont {Anasori}},
  \bibinfo {author} {\bibfnamefont {Y.}~\bibnamefont {Gogotsi}}, \ and\
  \bibinfo {author} {\bibfnamefont {Z.~W.}\ \bibnamefont {Seh}},\ }\bibfield
  {title} {\enquote {\bibinfo {title} {\textcolor{black}{Fundamentals of MXene
  synthesis}},}\ }\href {\doibase 10.1038/s44160-022-00104-6} {\bibfield
  {journal} {\bibinfo  {journal} {Nature Synthesis}\ }\textbf {\bibinfo
  {volume} {1}},\ \bibinfo {pages} {601--614} (\bibinfo {year}
  {2022})}\BibitemShut {NoStop}%
\bibitem [{\citenamefont {Ketolainen}\ and\ \citenamefont
  {Karlick{\'y}}(2022)}]{Ketolainen2022}%
  \BibitemOpen
  \bibfield  {author} {\bibinfo {author} {\bibfnamefont {T.}~\bibnamefont
  {Ketolainen}}\ and\ \bibinfo {author} {\bibfnamefont {F.}~\bibnamefont
  {Karlick{\'y}}},\ }\bibfield  {title} {\enquote {\bibinfo {title} {Optical
  gaps and excitons in semiconducting transition metal carbides (mxenes)},}\
  }\href {\doibase 10.1039/D2TC00246A} {\bibfield  {journal} {\bibinfo
  {journal} {J. Mater. Chem. C}\ }\textbf {\bibinfo {volume} {10}},\ \bibinfo
  {pages} {3919--3928} (\bibinfo {year} {2022})}\BibitemShut {NoStop}%
\bibitem [{\citenamefont {Sakhraoui}\ and\ \citenamefont
  {Karlick{\'y}}(2022)}]{Sakhraoui2022}%
  \BibitemOpen
  \bibfield  {author} {\bibinfo {author} {\bibfnamefont {T.}~\bibnamefont
  {Sakhraoui}}\ and\ \bibinfo {author} {\bibfnamefont {F.}~\bibnamefont
  {Karlick{\'y}}},\ }\bibfield  {title} {\enquote {\bibinfo {title} {Electronic
  nature transition and magnetism creation in vacancy-defected ti2co2 mxene
  under biaxial strain: A dftb + u study},}\ }\href {\doibase
  10.1021/acsomega.2c05037} {\bibfield  {journal} {\bibinfo  {journal} {ACS
  Omega}\ }\textbf {\bibinfo {volume} {7}},\ \bibinfo {pages} {42221--42232}
  (\bibinfo {year} {2022})}\BibitemShut {NoStop}%
\bibitem [{\citenamefont {Champagne}\ and\ \citenamefont
  {Charlier}(2020)}]{Champagne2021}%
  \BibitemOpen
  \bibfield  {author} {\bibinfo {author} {\bibfnamefont {A.}~\bibnamefont
  {Champagne}}\ and\ \bibinfo {author} {\bibfnamefont {J.-C.}\ \bibnamefont
  {Charlier}},\ }\bibfield  {title} {\enquote {\bibinfo {title} {Physical
  properties of 2d mxenes: from a theoretical perspective},}\ }\href {\doibase
  10.1088/2515-7639/ab97ee} {\bibfield  {journal} {\bibinfo  {journal} {J.
  Phys.: Materials}\ }\textbf {\bibinfo {volume} {3}},\ \bibinfo {pages}
  {032006} (\bibinfo {year} {2020})}\BibitemShut {NoStop}%
\bibitem [{\citenamefont {Kumar}\ and\ \citenamefont
  {Karlický}(2023)}]{Kumar2023}%
  \BibitemOpen
  \bibfield  {author} {\bibinfo {author} {\bibfnamefont {N.}~\bibnamefont
  {Kumar}}\ and\ \bibinfo {author} {\bibfnamefont {F.}~\bibnamefont
  {Karlický}},\ }\bibfield  {title} {\enquote {\bibinfo {title}
  {{Oxygen-terminated Ti3C2 MXene as an excitonic insulator}},}\ }\href
  {\doibase 10.1063/5.0143313} {\bibfield  {journal} {\bibinfo  {journal}
  {Applied Physics Letters}\ }\textbf {\bibinfo {volume} {122}},\ \bibinfo
  {pages} {183102} (\bibinfo {year} {2023})}\BibitemShut {NoStop}%
\bibitem [{\citenamefont {Lipatov}\ \emph {et~al.}(2018)\citenamefont
  {Lipatov}, \citenamefont {Lu}, \citenamefont {Alhabeb}, \citenamefont
  {Anasori}, \citenamefont {Gruverman}, \citenamefont {Gogotsi},\ and\
  \citenamefont {Sinitskii}}]{Lipatov2018}%
  \BibitemOpen
  \bibfield  {author} {\bibinfo {author} {\bibfnamefont {A.}~\bibnamefont
  {Lipatov}}, \bibinfo {author} {\bibfnamefont {H.}~\bibnamefont {Lu}},
  \bibinfo {author} {\bibfnamefont {M.}~\bibnamefont {Alhabeb}}, \bibinfo
  {author} {\bibfnamefont {B.}~\bibnamefont {Anasori}}, \bibinfo {author}
  {\bibfnamefont {A.}~\bibnamefont {Gruverman}}, \bibinfo {author}
  {\bibfnamefont {Y.}~\bibnamefont {Gogotsi}}, \ and\ \bibinfo {author}
  {\bibfnamefont {A.}~\bibnamefont {Sinitskii}},\ }\bibfield  {title} {\enquote
  {\bibinfo {title} {Elastic properties of 2d ti3c2tx mxene monolayers and
  bilayers},}\ }\href {\doibase 10.1126/sciadv.aat0491} {\bibfield  {journal}
  {\bibinfo  {journal} {Sci. Adv.}\ }\textbf {\bibinfo {volume} {4}},\ \bibinfo
  {pages} {eaat0491} (\bibinfo {year} {2018})}\BibitemShut {NoStop}%
\bibitem [{\citenamefont {Wyatt}, \citenamefont {Rosenkranz},\ and\
  \citenamefont {Anasori}(2021)}]{Wyatt2021}%
  \BibitemOpen
  \bibfield  {author} {\bibinfo {author} {\bibfnamefont {B.~C.}\ \bibnamefont
  {Wyatt}}, \bibinfo {author} {\bibfnamefont {A.}~\bibnamefont {Rosenkranz}}, \
  and\ \bibinfo {author} {\bibfnamefont {B.}~\bibnamefont {Anasori}},\
  }\bibfield  {title} {\enquote {\bibinfo {title} {2d mxenes: Tunable
  mechanical and tribological properties},}\ }\href {\doibase
  https://doi.org/10.1002/adma.202007973} {\bibfield  {journal} {\bibinfo
  {journal} {Adv. Mater.}\ }\textbf {\bibinfo {volume} {33}},\ \bibinfo {pages}
  {2007973} (\bibinfo {year} {2021})}\BibitemShut {NoStop}%
\bibitem [{\citenamefont {Khazaei}\ \emph {et~al.}(2012)\citenamefont
  {Khazaei}, \citenamefont {Arai}, \citenamefont {Sasaki}, \citenamefont
  {Chung}, \citenamefont {Venkataramanan}, \citenamefont {Estili},
  \citenamefont {Sakka},\ and\ \citenamefont {Kawazoe}}]{Khazaei2012}%
  \BibitemOpen
  \bibfield  {author} {\bibinfo {author} {\bibfnamefont {M.}~\bibnamefont
  {Khazaei}}, \bibinfo {author} {\bibfnamefont {M.}~\bibnamefont {Arai}},
  \bibinfo {author} {\bibfnamefont {T.}~\bibnamefont {Sasaki}}, \bibinfo
  {author} {\bibfnamefont {C.}~\bibnamefont {Chung}}, \bibinfo {author}
  {\bibfnamefont {N.~S.}\ \bibnamefont {Venkataramanan}}, \bibinfo {author}
  {\bibfnamefont {M.}~\bibnamefont {Estili}}, \bibinfo {author} {\bibfnamefont
  {Y.}~\bibnamefont {Sakka}}, \ and\ \bibinfo {author} {\bibfnamefont
  {Y.}~\bibnamefont {Kawazoe}},\ }\bibfield  {title} {\enquote {\bibinfo
  {title} {Novel electronic and magnetic properties of two‐dimensional
  transition metal carbides and nitrides},}\ }\href {\doibase
  10.1002/adfm.201202502} {\bibfield  {journal} {\bibinfo  {journal} {Adv.
  Funct. Mater.}\ }\textbf {\bibinfo {volume} {23}},\ \bibinfo {pages}
  {2185–2192} (\bibinfo {year} {2012})}\BibitemShut {NoStop}%
\bibitem [{\citenamefont {Thomas}\ and\ \citenamefont
  {Cherusseri}(2023)}]{Thomas2023}%
  \BibitemOpen
  \bibfield  {author} {\bibinfo {author} {\bibfnamefont {S.~A.}\ \bibnamefont
  {Thomas}}\ and\ \bibinfo {author} {\bibfnamefont {J.}~\bibnamefont
  {Cherusseri}},\ }\bibfield  {title} {\enquote {\bibinfo {title} {Recent
  advances in synthesis and properties of zirconium‐based mxenes for
  application in rechargeable batteries},}\ }\href@noop {} {\bibfield
  {journal} {\bibinfo  {journal} {Energy Storage}\ }\textbf {\bibinfo {volume}
  {5}} (\bibinfo {year} {2023})}\BibitemShut {NoStop}%
\bibitem [{\citenamefont {Duan}, \citenamefont {Zhou},\ and\ \citenamefont
  {Wang}(2023)}]{Duan2023}%
  \BibitemOpen
  \bibfield  {author} {\bibinfo {author} {\bibfnamefont {X.}~\bibnamefont
  {Duan}}, \bibinfo {author} {\bibfnamefont {B.}~\bibnamefont {Zhou}}, \ and\
  \bibinfo {author} {\bibfnamefont {X.}~\bibnamefont {Wang}},\ }\bibfield
  {title} {\enquote {\bibinfo {title} {Strain tailored electronic structure and
  magnetic properties of fe-doped zr8c4t8 (t = f, o) monolayers},}\ }\href
  {\doibase 10.1016/j.physe.2022.115488} {\bibfield  {journal} {\bibinfo
  {journal} {Phys. E: Low-Dimens.}\ }\textbf {\bibinfo {volume} {145}},\
  \bibinfo {pages} {115488} (\bibinfo {year} {2023})}\BibitemShut {NoStop}%
\bibitem [{\citenamefont {Zhang}\ \emph {et~al.}(2024)\citenamefont {Zhang},
  \citenamefont {Li}, \citenamefont {Zhang},\ and\ \citenamefont
  {Cui}}]{Zhang2024}%
  \BibitemOpen
  \bibfield  {author} {\bibinfo {author} {\bibfnamefont {H.}~\bibnamefont
  {Zhang}}, \bibinfo {author} {\bibfnamefont {X.-H.}\ \bibnamefont {Li}},
  \bibinfo {author} {\bibfnamefont {R.-Z.}\ \bibnamefont {Zhang}}, \ and\
  \bibinfo {author} {\bibfnamefont {H.-L.}\ \bibnamefont {Cui}},\ }\bibfield
  {title} {\enquote {\bibinfo {title} {First-principle study of electronic,
  optical, quantum capacitance, carrier mobility and photocatalytic properties
  of zr2co2 mxene under uniaxial strain},}\ }\href {\doibase
  10.1016/j.vacuum.2023.112813} {\bibfield  {journal} {\bibinfo  {journal}
  {Vacuum}\ }\textbf {\bibinfo {volume} {220}},\ \bibinfo {pages} {112813}
  (\bibinfo {year} {2024})}\BibitemShut {NoStop}%
\bibitem [{\citenamefont {Zha}\ \emph {et~al.}(2015)\citenamefont {Zha},
  \citenamefont {Luo}, \citenamefont {Li}, \citenamefont {Huang}, \citenamefont
  {He}, \citenamefont {Wen},\ and\ \citenamefont {Du}}]{Zha2015}%
  \BibitemOpen
  \bibfield  {author} {\bibinfo {author} {\bibfnamefont {X.-H.}\ \bibnamefont
  {Zha}}, \bibinfo {author} {\bibfnamefont {K.}~\bibnamefont {Luo}}, \bibinfo
  {author} {\bibfnamefont {Q.}~\bibnamefont {Li}}, \bibinfo {author}
  {\bibfnamefont {Q.}~\bibnamefont {Huang}}, \bibinfo {author} {\bibfnamefont
  {J.}~\bibnamefont {He}}, \bibinfo {author} {\bibfnamefont {X.}~\bibnamefont
  {Wen}}, \ and\ \bibinfo {author} {\bibfnamefont {S.}~\bibnamefont {Du}},\
  }\bibfield  {title} {\enquote {\bibinfo {title} {Role of the surface effect
  on the structural, electronic and mechanical properties of the carbide
  mxenes},}\ }\href {\doibase 10.1209/0295-5075/111/26007} {\bibfield
  {journal} {\bibinfo  {journal} {EPL}\ }\textbf {\bibinfo {volume} {111}},\
  \bibinfo {pages} {26007} (\bibinfo {year} {2015})}\BibitemShut {NoStop}%
\bibitem [{\citenamefont {Zhou}\ \emph {et~al.}(2016)\citenamefont {Zhou},
  \citenamefont {Zha}, \citenamefont {Chen}, \citenamefont {Ye}, \citenamefont
  {Eklund}, \citenamefont {Du},\ and\ \citenamefont {Huang}}]{Zhou2016}%
  \BibitemOpen
  \bibfield  {author} {\bibinfo {author} {\bibfnamefont {J.}~\bibnamefont
  {Zhou}}, \bibinfo {author} {\bibfnamefont {X.}~\bibnamefont {Zha}}, \bibinfo
  {author} {\bibfnamefont {F.~Y.}\ \bibnamefont {Chen}}, \bibinfo {author}
  {\bibfnamefont {Q.}~\bibnamefont {Ye}}, \bibinfo {author} {\bibfnamefont
  {P.}~\bibnamefont {Eklund}}, \bibinfo {author} {\bibfnamefont
  {S.}~\bibnamefont {Du}}, \ and\ \bibinfo {author} {\bibfnamefont
  {Q.}~\bibnamefont {Huang}},\ }\bibfield  {title} {\enquote {\bibinfo {title}
  {A two-dimensional zirconium carbide by selective etching of al3c3 from
  nanolaminated zr3al3c5},}\ }\href {\doibase
  https://doi.org/10.1002/anie.201510432} {\bibfield  {journal} {\bibinfo
  {journal} {Angew. Chem.-Int. Edit.}\ }\textbf {\bibinfo {volume} {55}},\
  \bibinfo {pages} {5008--5013} (\bibinfo {year} {2016})}\BibitemShut {NoStop}%
\bibitem [{\citenamefont {Zha}\ \emph {et~al.}(2022)\citenamefont {Zha},
  \citenamefont {Ma}, \citenamefont {Luo},\ and\ \citenamefont {Fu}}]{Zha2022}%
  \BibitemOpen
  \bibfield  {author} {\bibinfo {author} {\bibfnamefont {X.-H.}\ \bibnamefont
  {Zha}}, \bibinfo {author} {\bibfnamefont {X.}~\bibnamefont {Ma}}, \bibinfo
  {author} {\bibfnamefont {J.-T.}\ \bibnamefont {Luo}}, \ and\ \bibinfo
  {author} {\bibfnamefont {C.}~\bibnamefont {Fu}},\ }\bibfield  {title}
  {\enquote {\bibinfo {title} {Surface potential-determined performance of
  ti3c2t2 (t = o{,} f{,} oh) and zr3c2t2 (t = o{,} f{,} oh{,} s) mxenes as
  anode materials of sodium ion batteries},}\ }\href {\doibase
  10.1039/D2NR02271K} {\bibfield  {journal} {\bibinfo  {journal} {Nanoscale}\
  }\textbf {\bibinfo {volume} {14}},\ \bibinfo {pages} {10549--10558} (\bibinfo
  {year} {2022})}\BibitemShut {NoStop}%
\bibitem [{\citenamefont {Yang}\ \emph {et~al.}(2016)\citenamefont {Yang},
  \citenamefont {Peng}, \citenamefont {Sun},\ and\ \citenamefont
  {Perdew}}]{Yang2016}%
  \BibitemOpen
  \bibfield  {author} {\bibinfo {author} {\bibfnamefont {Z.}~\bibnamefont
  {Yang}}, \bibinfo {author} {\bibfnamefont {H.}~\bibnamefont {Peng}}, \bibinfo
  {author} {\bibfnamefont {J.}~\bibnamefont {Sun}}, \ and\ \bibinfo {author}
  {\bibfnamefont {J.~P.}\ \bibnamefont {Perdew}},\ }\bibfield  {title}
  {\enquote {\bibinfo {title} {More realistic band gaps from meta-generalized
  gradient approximations: Only in a generalized kohn-sham scheme},}\ }\href
  {\doibase 10.1103/PhysRevB.93.205205} {\bibfield  {journal} {\bibinfo
  {journal} {Phys. Rev. B}\ }\textbf {\bibinfo {volume} {93}},\ \bibinfo
  {pages} {205205} (\bibinfo {year} {2016})}\BibitemShut {NoStop}%
\bibitem [{\citenamefont {Khan}\ \emph {et~al.}(2017)\citenamefont {Khan},
  \citenamefont {Amin}, \citenamefont {Gan},\ and\ \citenamefont
  {Ahmad}}]{Khan2017}%
  \BibitemOpen
  \bibfield  {author} {\bibinfo {author} {\bibfnamefont {S.~A.}\ \bibnamefont
  {Khan}}, \bibinfo {author} {\bibfnamefont {B.}~\bibnamefont {Amin}}, \bibinfo
  {author} {\bibfnamefont {L.-Y.}\ \bibnamefont {Gan}}, \ and\ \bibinfo
  {author} {\bibfnamefont {I.}~\bibnamefont {Ahmad}},\ }\bibfield  {title}
  {\enquote {\bibinfo {title} {Strain engineering of electronic structures and
  photocatalytic responses of mxenes functionalized by oxygen},}\ }\href
  {\doibase 10.1039/c7cp02513k} {\bibfield  {journal} {\bibinfo  {journal}
  {Phys. Chem. Chem. Phys.}\ }\textbf {\bibinfo {volume} {19}},\ \bibinfo
  {pages} {14738–14744} (\bibinfo {year} {2017})}\BibitemShut {NoStop}%
\bibitem [{\citenamefont {Cui}\ \emph {et~al.}(2019)\citenamefont {Cui},
  \citenamefont {Peng}, \citenamefont {Zhou},\ and\ \citenamefont
  {Sun}}]{Cui2019}%
  \BibitemOpen
  \bibfield  {author} {\bibinfo {author} {\bibfnamefont {J.}~\bibnamefont
  {Cui}}, \bibinfo {author} {\bibfnamefont {Q.}~\bibnamefont {Peng}}, \bibinfo
  {author} {\bibfnamefont {J.}~\bibnamefont {Zhou}}, \ and\ \bibinfo {author}
  {\bibfnamefont {Z.}~\bibnamefont {Sun}},\ }\bibfield  {title} {\enquote
  {\bibinfo {title} {Strain-tunable electronic structures and optical
  properties of semiconducting mxenes},}\ }\href {\doibase
  10.1088/1361-6528/ab1f22} {\bibfield  {journal} {\bibinfo  {journal}
  {Nanotechnology}\ }\textbf {\bibinfo {volume} {30}},\ \bibinfo {pages}
  {345205} (\bibinfo {year} {2019})}\BibitemShut {NoStop}%
\bibitem [{\citenamefont {Zhang}\ \emph {et~al.}(2020)\citenamefont {Zhang},
  \citenamefont {Zha}, \citenamefont {Luo}, \citenamefont {Qin}, \citenamefont
  {Bai}, \citenamefont {Xu}, \citenamefont {Lin}, \citenamefont {Huang},\ and\
  \citenamefont {Du}}]{Zhang2020}%
  \BibitemOpen
  \bibfield  {author} {\bibinfo {author} {\bibfnamefont {Y.}~\bibnamefont
  {Zhang}}, \bibinfo {author} {\bibfnamefont {X.-H.}\ \bibnamefont {Zha}},
  \bibinfo {author} {\bibfnamefont {K.}~\bibnamefont {Luo}}, \bibinfo {author}
  {\bibfnamefont {Y.}~\bibnamefont {Qin}}, \bibinfo {author} {\bibfnamefont
  {X.}~\bibnamefont {Bai}}, \bibinfo {author} {\bibfnamefont {J.}~\bibnamefont
  {Xu}}, \bibinfo {author} {\bibfnamefont {C.-T.}\ \bibnamefont {Lin}},
  \bibinfo {author} {\bibfnamefont {Q.}~\bibnamefont {Huang}}, \ and\ \bibinfo
  {author} {\bibfnamefont {S.}~\bibnamefont {Du}},\ }\bibfield  {title}
  {\enquote {\bibinfo {title} {Theoretical study on the electrical and
  mechanical properties of mxene multilayer structures through strain
  regulation},}\ }\href {\doibase https://doi.org/10.1016/j.cplett.2020.137997}
  {\bibfield  {journal} {\bibinfo  {journal} {Chem. Phys. Lett.}\ }\textbf
  {\bibinfo {volume} {760}},\ \bibinfo {pages} {137997} (\bibinfo {year}
  {2020})}\BibitemShut {NoStop}%
\bibitem [{\citenamefont {Hedin}(1965)}]{Hedin1965}%
  \BibitemOpen
  \bibfield  {author} {\bibinfo {author} {\bibfnamefont {L.}~\bibnamefont
  {Hedin}},\ }\bibfield  {title} {\enquote {\bibinfo {title} {New method for
  calculating the one-particle {Green’s} function with application to the
  electron-gas problem},}\ }\href {\doibase 10.1103/PhysRev.139.A796}
  {\bibfield  {journal} {\bibinfo  {journal} {Phys. Rev.}\ }\textbf {\bibinfo
  {volume} {139}},\ \bibinfo {pages} {A796--A823} (\bibinfo {year}
  {1965})}\BibitemShut {NoStop}%
\bibitem [{\citenamefont {Bethe}\ and\ \citenamefont
  {Salpeter}(1951)}]{Bethe1951}%
  \BibitemOpen
  \bibfield  {author} {\bibinfo {author} {\bibfnamefont {H.}~\bibnamefont
  {Bethe}}\ and\ \bibinfo {author} {\bibfnamefont {E.}~\bibnamefont
  {Salpeter}},\ }\bibfield  {title} {\enquote {\bibinfo {title} {A relativistic
  equation for bound state problems},}\ }\href@noop {} {\bibfield  {journal}
  {\bibinfo  {journal} {Phys. Rev.}\ }\textbf {\bibinfo {volume} {82}},\
  \bibinfo {pages} {309--310} (\bibinfo {year} {1951})}\BibitemShut {NoStop}%
\bibitem [{\citenamefont {Kresse}\ and\ \citenamefont
  {Hafner}(1993)}]{kresse_ab_1993}%
  \BibitemOpen
  \bibfield  {author} {\bibinfo {author} {\bibfnamefont {G.}~\bibnamefont
  {Kresse}}\ and\ \bibinfo {author} {\bibfnamefont {J.}~\bibnamefont
  {Hafner}},\ }\bibfield  {title} {\enquote {\bibinfo {title} {Ab initio
  molecular dynamics for open-shell transition metals},}\ }\href {\doibase
  10.1103/physrevb.48.13115} {\bibfield  {journal} {\bibinfo  {journal} {Phys.
  Rev. B}\ }\textbf {\bibinfo {volume} {48}},\ \bibinfo {pages} {13115--13118}
  (\bibinfo {year} {1993})}\BibitemShut {NoStop}%
\bibitem [{\citenamefont {Kresse}\ and\ \citenamefont
  {Furthm\"{u}ller}(1996{\natexlab{a}})}]{kresse_efficiency_1996}%
  \BibitemOpen
  \bibfield  {author} {\bibinfo {author} {\bibfnamefont {G.}~\bibnamefont
  {Kresse}}\ and\ \bibinfo {author} {\bibfnamefont {J.}~\bibnamefont
  {Furthm\"{u}ller}},\ }\bibfield  {title} {\enquote {\bibinfo {title}
  {Efficiency of ab-initio total energy calculations for metals and
  semiconductors using a plane-wave basis set},}\ }\href {\doibase
  10.1016/0927-0256(96)00008-0} {\bibfield  {journal} {\bibinfo  {journal}
  {Comput. Mater. Sci.}\ }\textbf {\bibinfo {volume} {6}},\ \bibinfo {pages}
  {15--50} (\bibinfo {year} {1996}{\natexlab{a}})}\BibitemShut {NoStop}%
\bibitem [{\citenamefont {Kresse}\ and\ \citenamefont
  {Furthm\"{u}ller}(1996{\natexlab{b}})}]{kresse_efficient_1996}%
  \BibitemOpen
  \bibfield  {author} {\bibinfo {author} {\bibfnamefont {G.}~\bibnamefont
  {Kresse}}\ and\ \bibinfo {author} {\bibfnamefont {J.}~\bibnamefont
  {Furthm\"{u}ller}},\ }\bibfield  {title} {\enquote {\bibinfo {title}
  {Efficient iterative schemes forab initiototal-energy calculations using a
  plane-wave basis set},}\ }\href {\doibase 10.1103/physrevb.54.11169}
  {\bibfield  {journal} {\bibinfo  {journal} {Phys. Rev. B}\ }\textbf {\bibinfo
  {volume} {54}},\ \bibinfo {pages} {11169--11186} (\bibinfo {year}
  {1996}{\natexlab{b}})}\BibitemShut {NoStop}%
\bibitem [{\citenamefont {Kresse}\ and\ \citenamefont
  {Hafner}(1994)}]{kresse_norm-conserving_1994}%
  \BibitemOpen
  \bibfield  {author} {\bibinfo {author} {\bibfnamefont {G.}~\bibnamefont
  {Kresse}}\ and\ \bibinfo {author} {\bibfnamefont {J.}~\bibnamefont
  {Hafner}},\ }\bibfield  {title} {\enquote {\bibinfo {title} {Norm-conserving
  and ultrasoft pseudopotentials for first-row and transition elements},}\
  }\href {\doibase 10.1088/0953-8984/6/40/015} {\bibfield  {journal} {\bibinfo
  {journal} {J. Phys.: Condens. Matter}\ }\textbf {\bibinfo {volume} {6}},\
  \bibinfo {pages} {8245--8257} (\bibinfo {year} {1994})}\BibitemShut {NoStop}%
\bibitem [{\citenamefont {Bl\"{o}chl}(1994)}]{Blochl1994}%
  \BibitemOpen
  \bibfield  {author} {\bibinfo {author} {\bibfnamefont {P.~E.}\ \bibnamefont
  {Bl\"{o}chl}},\ }\bibfield  {title} {\enquote {\bibinfo {title} {Projector
  augmented-wave method},}\ }\href {\doibase 10.1103/physrevb.50.17953}
  {\bibfield  {journal} {\bibinfo  {journal} {Phys. Rev. B}\ }\textbf {\bibinfo
  {volume} {50}},\ \bibinfo {pages} {17953--17979} (\bibinfo {year}
  {1994})}\BibitemShut {NoStop}%
\bibitem [{\citenamefont {Perdew}, \citenamefont {Burke},\ and\ \citenamefont
  {Ernzerhof}(1996)}]{PBE}%
  \BibitemOpen
  \bibfield  {author} {\bibinfo {author} {\bibfnamefont {J.~P.}\ \bibnamefont
  {Perdew}}, \bibinfo {author} {\bibfnamefont {K.}~\bibnamefont {Burke}}, \
  and\ \bibinfo {author} {\bibfnamefont {M.}~\bibnamefont {Ernzerhof}},\
  }\bibfield  {title} {\enquote {\bibinfo {title} {Generalized gradient
  approximation made simple},}\ }\href {\doibase 10.1103/physrevlett.77.3865}
  {\bibfield  {journal} {\bibinfo  {journal} {Phys. Rev. Lett.}\ }\textbf
  {\bibinfo {volume} {77}},\ \bibinfo {pages} {3865--3868} (\bibinfo {year}
  {1996})}\BibitemShut {NoStop}%
\bibitem [{\citenamefont {Sun}, \citenamefont {Ruzsinszky},\ and\ \citenamefont
  {Perdew}(2015)}]{SCAN}%
  \BibitemOpen
  \bibfield  {author} {\bibinfo {author} {\bibfnamefont {J.}~\bibnamefont
  {Sun}}, \bibinfo {author} {\bibfnamefont {A.}~\bibnamefont {Ruzsinszky}}, \
  and\ \bibinfo {author} {\bibfnamefont {J.~P.}\ \bibnamefont {Perdew}},\
  }\bibfield  {title} {\enquote {\bibinfo {title} {Strongly constrained and
  appropriately normed semilocal density functional},}\ }\href {\doibase
  10.1103/PhysRevLett.115.036402} {\bibfield  {journal} {\bibinfo  {journal}
  {Phys. Rev. Lett.}\ }\textbf {\bibinfo {volume} {115}},\ \bibinfo {pages}
  {036402} (\bibinfo {year} {2015})}\BibitemShut {NoStop}%
\bibitem [{\citenamefont {Krukau}\ \emph {et~al.}(2006)\citenamefont {Krukau},
  \citenamefont {Vydrov}, \citenamefont {Izmaylov},\ and\ \citenamefont
  {Scuseria}}]{HSE06}%
  \BibitemOpen
  \bibfield  {author} {\bibinfo {author} {\bibfnamefont {A.~V.}\ \bibnamefont
  {Krukau}}, \bibinfo {author} {\bibfnamefont {O.~A.}\ \bibnamefont {Vydrov}},
  \bibinfo {author} {\bibfnamefont {A.~F.}\ \bibnamefont {Izmaylov}}, \ and\
  \bibinfo {author} {\bibfnamefont {G.~E.}\ \bibnamefont {Scuseria}},\
  }\bibfield  {title} {\enquote {\bibinfo {title} {Influence of the exchange
  screening parameter on the performance of screened hybrid functionals},}\
  }\href {\doibase 10.1063/1.2404663} {\bibfield  {journal} {\bibinfo
  {journal} {J. Chem. Phys.}\ }\textbf {\bibinfo {volume} {125}} (\bibinfo
  {year} {2006}),\ 10.1063/1.2404663}\BibitemShut {NoStop}%
\bibitem [{\citenamefont {Togo}\ and\ \citenamefont {Tanaka}(2015)}]{Togo2015}%
  \BibitemOpen
  \bibfield  {author} {\bibinfo {author} {\bibfnamefont {A.}~\bibnamefont
  {Togo}}\ and\ \bibinfo {author} {\bibfnamefont {I.}~\bibnamefont {Tanaka}},\
  }\bibfield  {title} {\enquote {\bibinfo {title} {First principles phonon
  calculations in materials science},}\ }\href {\doibase
  10.1016/j.scriptamat.2015.07.021} {\bibfield  {journal} {\bibinfo  {journal}
  {Scr. Mater.}\ }\textbf {\bibinfo {volume} {108}},\ \bibinfo {pages} {1--5}
  (\bibinfo {year} {2015})}\BibitemShut {NoStop}%
\bibitem [{\citenamefont {Shishkin}\ and\ \citenamefont
  {Kresse}(2006)}]{Shishkin2006}%
  \BibitemOpen
  \bibfield  {author} {\bibinfo {author} {\bibfnamefont {M.}~\bibnamefont
  {Shishkin}}\ and\ \bibinfo {author} {\bibfnamefont {G.}~\bibnamefont
  {Kresse}},\ }\bibfield  {title} {\enquote {\bibinfo {title} {Implementation
  and performance of the frequency-dependent $gw$ method within the paw
  framework},}\ }\href {\doibase 10.1103/PhysRevB.74.035101} {\bibfield
  {journal} {\bibinfo  {journal} {Phys. Rev. B}\ }\textbf {\bibinfo {volume}
  {74}},\ \bibinfo {pages} {035101} (\bibinfo {year} {2006})}\BibitemShut
  {NoStop}%
\bibitem [{\citenamefont {Kolos}\ and\ \citenamefont
  {Karlick\'{y}}(2019)}]{Kolos2019}%
  \BibitemOpen
  \bibfield  {author} {\bibinfo {author} {\bibfnamefont {M.}~\bibnamefont
  {Kolos}}\ and\ \bibinfo {author} {\bibfnamefont {F.}~\bibnamefont
  {Karlick\'{y}}},\ }\bibfield  {title} {\enquote {\bibinfo {title} {Accurate
  many-body calculation of electronic and optical band gap of bulk hexagonal
  boron nitride},}\ }\href {\doibase 10.1039/C8CP07328G} {\bibfield  {journal}
  {\bibinfo  {journal} {Phys. Chem. Chem. Phys.}\ }\textbf {\bibinfo {volume}
  {21}},\ \bibinfo {pages} {3999--4005} (\bibinfo {year} {2019})}\BibitemShut
  {NoStop}%
\bibitem [{\citenamefont {Dubeck{\'{y}}}\ \emph {et~al.}(2020)\citenamefont
  {Dubeck{\'{y}}}, \citenamefont {Karlick{\'{y}}}, \citenamefont
  {Min{\'{a}}rik},\ and\ \citenamefont {Mitas}}]{Dubecky2020}%
  \BibitemOpen
  \bibfield  {author} {\bibinfo {author} {\bibfnamefont {M.}~\bibnamefont
  {Dubeck{\'{y}}}}, \bibinfo {author} {\bibfnamefont {F.}~\bibnamefont
  {Karlick{\'{y}}}}, \bibinfo {author} {\bibfnamefont {S.}~\bibnamefont
  {Min{\'{a}}rik}}, \ and\ \bibinfo {author} {\bibfnamefont {L.}~\bibnamefont
  {Mitas}},\ }\bibfield  {title} {\enquote {\bibinfo {title} {Fundamental gap
  of fluorographene by many-body {GW} and fixed-node diffusion monte carlo
  methods},}\ }\href {\doibase 10.1063/5.0030952} {\bibfield  {journal}
  {\bibinfo  {journal} {J. Chem. Phys.}\ }\textbf {\bibinfo {volume} {153}},\
  \bibinfo {pages} {184706} (\bibinfo {year} {2020})}\BibitemShut {NoStop}%
\bibitem [{\citenamefont {Kolos}\ and\ \citenamefont
  {Karlick{\'{y}}}(2022)}]{Kolos2022}%
  \BibitemOpen
  \bibfield  {author} {\bibinfo {author} {\bibfnamefont {M.}~\bibnamefont
  {Kolos}}\ and\ \bibinfo {author} {\bibfnamefont {F.}~\bibnamefont
  {Karlick{\'{y}}}},\ }\bibfield  {title} {\enquote {\bibinfo {title} {The
  electronic and optical properties of {III}{\textendash}v binary 2d
  semiconductors: how to achieve high precision from accurate many-body
  methods},}\ }\href {\doibase 10.1039/d2cp04432c} {\bibfield  {journal}
  {\bibinfo  {journal} {Phys. Chem. Chem. Phys.}\ }\textbf {\bibinfo {volume}
  {24}},\ \bibinfo {pages} {27459--27466} (\bibinfo {year} {2022})}\BibitemShut
  {NoStop}%
\bibitem [{\citenamefont {Albrecht}\ \emph {et~al.}(1998)\citenamefont
  {Albrecht}, \citenamefont {Reining}, \citenamefont {Del~Sole},\ and\
  \citenamefont {Onida}}]{Albrecht1998}%
  \BibitemOpen
  \bibfield  {author} {\bibinfo {author} {\bibfnamefont {S.}~\bibnamefont
  {Albrecht}}, \bibinfo {author} {\bibfnamefont {L.}~\bibnamefont {Reining}},
  \bibinfo {author} {\bibfnamefont {R.}~\bibnamefont {Del~Sole}}, \ and\
  \bibinfo {author} {\bibfnamefont {G.}~\bibnamefont {Onida}},\ }\bibfield
  {title} {\enquote {\bibinfo {title} {Ab initio calculation of excitonic
  effects in the optical spectra of semiconductors},}\ }\href {\doibase
  10.1103/PhysRevLett.80.4510} {\bibfield  {journal} {\bibinfo  {journal}
  {Phys. Rev. Lett.}\ }\textbf {\bibinfo {volume} {80}},\ \bibinfo {pages}
  {4510--4513} (\bibinfo {year} {1998})}\BibitemShut {NoStop}%
\bibitem [{\citenamefont {Karlick{\'{y}}}\ and\ \citenamefont
  {Turo{\v{n}}}(2018)}]{Karlicky2018}%
  \BibitemOpen
  \bibfield  {author} {\bibinfo {author} {\bibfnamefont {F.}~\bibnamefont
  {Karlick{\'{y}}}}\ and\ \bibinfo {author} {\bibfnamefont {J.}~\bibnamefont
  {Turo{\v{n}}}},\ }\bibfield  {title} {\enquote {\bibinfo {title}
  {Fluorographane c2fh: Stable and wide band gap insulator with huge excitonic
  effect},}\ }\href {\doibase 10.1016/j.carbon.2018.04.006} {\bibfield
  {journal} {\bibinfo  {journal} {Carbon}\ }\textbf {\bibinfo {volume} {135}},\
  \bibinfo {pages} {134--144} (\bibinfo {year} {2018})}\BibitemShut {NoStop}%
\bibitem [{\citenamefont {Dubecký}, \citenamefont {Minárik},\ and\
  \citenamefont {Karlický}(2023)}]{Dubecky2023}%
  \BibitemOpen
  \bibfield  {author} {\bibinfo {author} {\bibfnamefont {M.}~\bibnamefont
  {Dubecký}}, \bibinfo {author} {\bibfnamefont {S.}~\bibnamefont {Minárik}},
  \ and\ \bibinfo {author} {\bibfnamefont {F.}~\bibnamefont {Karlický}},\
  }\bibfield  {title} {\enquote {\bibinfo {title} {Benchmarking fundamental gap
  of sc2c(oh)2 mxene by many-body methods},}\ }\href
  {http://dx.doi.org/10.1063/5.0140315} {\bibfield  {journal} {\bibinfo
  {journal} {J. Chem. Phys.}\ }\textbf {\bibinfo {volume} {158}},\ \bibinfo
  {pages} {054703} (\bibinfo {year} {2023})}\BibitemShut {NoStop}%
\bibitem [{\citenamefont {Berseneva}\ \emph {et~al.}(2013)\citenamefont
  {Berseneva}, \citenamefont {Gulans}, \citenamefont {Krasheninnikov},\ and\
  \citenamefont {Nieminen}}]{Berseneva2013}%
  \BibitemOpen
  \bibfield  {author} {\bibinfo {author} {\bibfnamefont {N.}~\bibnamefont
  {Berseneva}}, \bibinfo {author} {\bibfnamefont {A.}~\bibnamefont {Gulans}},
  \bibinfo {author} {\bibfnamefont {A.~V.}\ \bibnamefont {Krasheninnikov}}, \
  and\ \bibinfo {author} {\bibfnamefont {R.~M.}\ \bibnamefont {Nieminen}},\
  }\bibfield  {title} {\enquote {\bibinfo {title} {Electronic structure of
  boron nitride sheets doped with carbon from first-principles calculations},}\
  }\href {\doibase 10.1103/physrevb.87.035404} {\bibfield  {journal} {\bibinfo
  {journal} {Phys. Rev. B}\ }\textbf {\bibinfo {volume} {87}},\ \bibinfo
  {pages} {035404} (\bibinfo {year} {2013})}\BibitemShut {NoStop}%
\bibitem [{\citenamefont {Choi}\ \emph {et~al.}(2015)\citenamefont {Choi},
  \citenamefont {Cui}, \citenamefont {Lan},\ and\ \citenamefont
  {Zhang}}]{Choi2015}%
  \BibitemOpen
  \bibfield  {author} {\bibinfo {author} {\bibfnamefont {J.-H.}\ \bibnamefont
  {Choi}}, \bibinfo {author} {\bibfnamefont {P.}~\bibnamefont {Cui}}, \bibinfo
  {author} {\bibfnamefont {H.}~\bibnamefont {Lan}}, \ and\ \bibinfo {author}
  {\bibfnamefont {Z.}~\bibnamefont {Zhang}},\ }\bibfield  {title} {\enquote
  {\bibinfo {title} {Linear scaling of the exciton binding energy versus the
  band gap of two-dimensional materials},}\ }\href {\doibase
  10.1103/physrevlett.115.066403} {\bibfield  {journal} {\bibinfo  {journal}
  {Phys. Rev. Lett.}\ }\textbf {\bibinfo {volume} {115}},\ \bibinfo {pages}
  {066403} (\bibinfo {year} {2015})}\BibitemShut {NoStop}%
\bibitem [{\citenamefont {Qian}\ \emph {et~al.}(2014)\citenamefont {Qian},
  \citenamefont {Liu}, \citenamefont {Fu},\ and\ \citenamefont
  {Li}}]{Qian2014}%
  \BibitemOpen
  \bibfield  {author} {\bibinfo {author} {\bibfnamefont {X.}~\bibnamefont
  {Qian}}, \bibinfo {author} {\bibfnamefont {J.}~\bibnamefont {Liu}}, \bibinfo
  {author} {\bibfnamefont {L.}~\bibnamefont {Fu}}, \ and\ \bibinfo {author}
  {\bibfnamefont {J.}~\bibnamefont {Li}},\ }\bibfield  {title} {\enquote
  {\bibinfo {title} {Quantum spin hall effect in two-dimensional transition
  metal dichalcogenides},}\ }\href@noop {} {\bibfield  {journal} {\bibinfo
  {journal} {Science}\ }\textbf {\bibinfo {volume} {346}},\ \bibinfo {pages}
  {1344--1347} (\bibinfo {year} {2014})}\BibitemShut {NoStop}%
\bibitem [{\citenamefont {Kumar}\ \emph {et~al.}(2024)\citenamefont {Kumar},
  \citenamefont {Kolos}, \citenamefont {Bhattacharya},\ and\ \citenamefont
  {Karlick\'{y}}}]{Kumar2024}%
  \BibitemOpen
  \bibfield  {author} {\bibinfo {author} {\bibfnamefont {N.}~\bibnamefont
  {Kumar}}, \bibinfo {author} {\bibfnamefont {M.}~\bibnamefont {Kolos}},
  \bibinfo {author} {\bibfnamefont {S.}~\bibnamefont {Bhattacharya}}, \ and\
  \bibinfo {author} {\bibfnamefont {F.}~\bibnamefont {Karlick\'{y}}},\
  }\bibfield  {title} {\enquote {\bibinfo {title} {Excitons, optical spectra,
  and electronic properties of semiconducting hf-based mxenes},}\ }\href
  {\doibase https://doi.org/10.1063/5.0197238} {\bibfield  {journal} {\bibinfo
  {journal} {J. Chem. Phys.}\ }\textbf {\bibinfo {volume} {160}},\ \bibinfo
  {pages} {124707} (\bibinfo {year} {2024})}\BibitemShut {NoStop}%
\bibitem [{\citenamefont {Ketolainen}, \citenamefont {Macháčová},\ and\
  \citenamefont {Karlický}(2020)}]{Ketolainen2020}%
  \BibitemOpen
  \bibfield  {author} {\bibinfo {author} {\bibfnamefont {T.}~\bibnamefont
  {Ketolainen}}, \bibinfo {author} {\bibfnamefont {N.}~\bibnamefont
  {Macháčová}}, \ and\ \bibinfo {author} {\bibfnamefont {F.}~\bibnamefont
  {Karlický}},\ }\bibfield  {title} {\enquote {\bibinfo {title} {Optical gaps
  and excitonic properties of 2d materials by hybrid time-dependent density
  functional theory: Evidences for monolayers and prospects for van der waals
  heterostructures},}\ }\href {\doibase 10.1021/acs.jctc.0c00387} {\bibfield
  {journal} {\bibinfo  {journal} {J. Chem. Theory Comput.}\ }\textbf {\bibinfo
  {volume} {16}},\ \bibinfo {pages} {5876–5883} (\bibinfo {year}
  {2020})}\BibitemShut {NoStop}%
\bibitem [{\citenamefont {Jiang}\ \emph {et~al.}(2017)\citenamefont {Jiang},
  \citenamefont {Liu}, \citenamefont {Li},\ and\ \citenamefont
  {Duan}}]{Jiang2017}%
  \BibitemOpen
  \bibfield  {author} {\bibinfo {author} {\bibfnamefont {Z.}~\bibnamefont
  {Jiang}}, \bibinfo {author} {\bibfnamefont {Z.}~\bibnamefont {Liu}}, \bibinfo
  {author} {\bibfnamefont {Y.}~\bibnamefont {Li}}, \ and\ \bibinfo {author}
  {\bibfnamefont {W.}~\bibnamefont {Duan}},\ }\bibfield  {title} {\enquote
  {\bibinfo {title} {Scaling universality between band gap and exciton binding
  energy of two-dimensional semiconductors},}\ }\href {\doibase
  10.1103/PhysRevLett.118.266401} {\bibfield  {journal} {\bibinfo  {journal}
  {Phys. Rev. Lett.}\ }\textbf {\bibinfo {volume} {118}},\ \bibinfo {pages}
  {266401} (\bibinfo {year} {2017})}\BibitemShut {NoStop}%
\end{thebibliography}%

\end{document}